\documentclass[english,aps,preprint]{revtex4-1}
\usepackage[T1]{fontenc}
\usepackage[latin9]{inputenc}
\usepackage{verbatim}
\usepackage{float}
\usepackage{units}
\usepackage{amsmath}
\usepackage{amssymb}
\usepackage{graphicx}
\usepackage[pdftex,urlcolor=blue,citecolor=blue,colorlinks=true,linkcolor=blue]{hyperref}
\usepackage{babel}
\begin{document}
\title{Folded chaotic whispering-gallery modes in non-convex, waveguide-coupled
planar optical microresonators}
\author{Kahli Burke and Jens U. N{\"o}ckel}
\email[To whom correspondence should be addressed: ]{noeckel@uoregon.edu}

\affiliation{Department of Physics \\
University of Oregon}
\date{\today}
\begin{abstract}
Chaotic whispering-gallery modes have significance both for optical
applications and for our understanding
of the interplay between wave phenomena and the classical ray limit
in the presence of chaotic dynamics and openness. In strongly non-convex
geometries, a theorem by Mather rules out the existence of invariant curves in phase
space corresponding to rays circulating in whispering-gallery patterns, so
that no corresponding modes of this type are expected.
Here we discuss numerical computations of the electromagnetic fields
in planar dielectric cavities that are strongly non-convex because
they are coupled to waveguides. We find a family of special states
which retains many features of the chaotic whispering-gallery modes
known from convex shapes: an intensity pattern corresponding to near-grazing
incidence along extended parts of the boundary, and comparatively high cavity Q factors.
The modes are folded into a figure-eight pattern, so overlap with the boundary is reduced
in the region of self-intersection. The modes combine the
phenomenology of chaotic whispering-gallery modes with an important technological
advantage: the ability to directly attach waveguides without spoiling
the Q factor of the folded mode. Using both a boundary-integral method and
the finite-difference time domain technique, we explore the dependence of
the phenomenon on wavelength in relation to cavity size, refractive-index contrast
to the surrounding medium, and the degree of shape deformation. A novel feature
that distinguishes folded from regular whispering gallery modes is
that a given shape will support high-Q folded chaotic whispering gallery modes only
in certain wavelength windows.
\end{abstract}

\maketitle
\begin{comment}
Data are on Talapas
\end{comment}

\section{Introduction}

Wave-chaotic optical cavities are of interest in applications because
of the degrees of design freedom that are opened up when the
geometric constraints of separability and symmetry are removed \citep{CaoWiersigReview}: spectral
density, lifetimes and emission directionality can all be tailored
in sometimes dramatic ways that go far beyond perturbative effects
around separable special cases. Among the first examples are planar
dielectric ovals that support \emph{chaotic whispering-gallery modes}
(WGMs) \citep{Nockel:1997lr}, where the anisotropic shape gives rise
to highly universal directional emission patterns that can be explained
based on the classical phase-space structure of the ray dynamics.
From a theoretical point of view, this connection to classical ray
physics is intriguing because the transition to chaos that occurs
in generic nonseparable systems must then be studied in the presence
of openness. The wave equation in this scenario exhibits quasibound
states at complex wavenumber $k=k_{\mathrm{r}}-i\,\kappa$; here $\kappa$
measures the escape rate which is related to the quality factor via
\begin{equation}
Q=\frac{k}{2\kappa}.\label{eq:QfactorDef}
\end{equation}
WGMs in oval resonators are special states that preserve long lifetimes
comparable to those found in circular dielectrics \citep{ISI:000320621600008},
due to the fact that they are predominantly localized in the phase-space
region corresponding to total internal reflection at the dielectric
boundary. When the invariant curves (tori) that foliate this phase-space
region in the limit of circular symmetry are gradually broken up following
the KAM theorem as applied to convex billiards \citep{LazutkinBook},
regular WGMs become \emph{chaotic} WGMs in which rays explore regions
of phase space in which the rays cease to be confined by total internal
reflection. The critical angle for total internal reflection defines
an escape window in phase space rather than real space, because this
angle is directly related to the tangential momentum component at
the surface of the resonator. In Ref.\  \citep{Nockel:1997lr}, an approximate
semiclassical quantization of these unconventional modes was proposed
based on a separation of time scales between the fast whispering-gallery
circulation and a slower spiraling-in toward the escape window, making
it possible to identify \emph{adiabatic invariant curves.} Their location
then served as an initial condition for ray simulations of the escape
directionality, caused by the mixed phase space in the vicinity of
the escape window.

It has long been understood that no true WGMs can be sustained in
circular domains if \emph{waveguides} are attached, because this creates
openings in the boundary that interrupt the whispering-gallery circulation.
In phase space, the escape window then depends not only on tangential
momentum but on position. For a detailed study of this breakdown in
the context of electronic microstructures with otherwise impenetrable
walls, see Ref.\  \citep{PhysRevB.54.10652}. This can be viewed as
a special case of a theorem by Mather \citep{MatherInvariant}, who
proved the non-existence of whispering-gallery invariant curves in
planar billiards when the boundary is not everywhere convex. An example
for Mather's theorem is the Bunimovich stadium (two semicircles joined
by straight, parallel sides) \citep{BunimovichStadium}, whose ray
dynamics does not permit whispering-gallery orbits even if $\ell$
is arbitrarily short, even though the weaker condition of non-concavity
still holds. Strict convexity is therefore required in order to sustain
WG circulation, but any waveguide openings will necessarily introduce
corners where this condition breaks down.

By exploiting the preferential emission directions of chaotic WGMs
in the near or far field, light can be coupled into and out of the
resonator without contacting (and thereby perturbing) the boundary
geometry directly. However, monolithically attached waveguides offer
some distinct advantages from an engineering point of view. In laser
applications, they allow both efficient optical pumping and more complete
collection of the emission. Although directional emission patterns
from planar cavities can be made highly directional \emph{in the plane}
 \citep{Nockel:94}, losses by out-of-plane diffraction can be significant
 \citep{backes:VerticalAndNotches,VerticalLossWithWG,VerticalSpreadingDisk,LimaconDirectionality}.
With attached waveguides, such losses are minimized. Moreover, waveguides
afford precise control over the number of input/output channels.

This has motivated several recent proposals to integrate chaotic cavities
with waveguides
 \citep{ChaoticChannelingPRL,EndFireInjection,MicrospiralWG,SquareMicrolasers,TriangleAndSquareWG,octagonWithLead}.
Because of Mather's theorem, however, almost all such proposals involve
modes that are \emph{not} of whispering-gallery type, achieving good
confinement instead with states localized on periodic orbits: Stable
and unstable periodic ray orbits whose reflections occur far enough
away from any waveguide openings will be insensitive to the boundary
shape at the waveguide apertures. These periodic orbits exist with
or without the presence of attached waveguides, and they make lasing
possible even in the free-standing Bunimovich stadium \citep{Fukushima:07}.

The appearance of wavefunction scarring due to unstable periodic orbits
in \citep{Fukushima:07} illustrates that wave solutions of the Helmholtz
equation can defy ray-optics predictions. This is also true for the
predictions of Mather's theorem, and again the stadium billiard is
a case in point: Numerical solutions for closed (hard-wall) cavities
 \citep{Prange2001} show that a form of chaotic WGM exists if the
straight sides are sufficiently short.
\begin{figure}
\includegraphics[width=0.9\columnwidth]{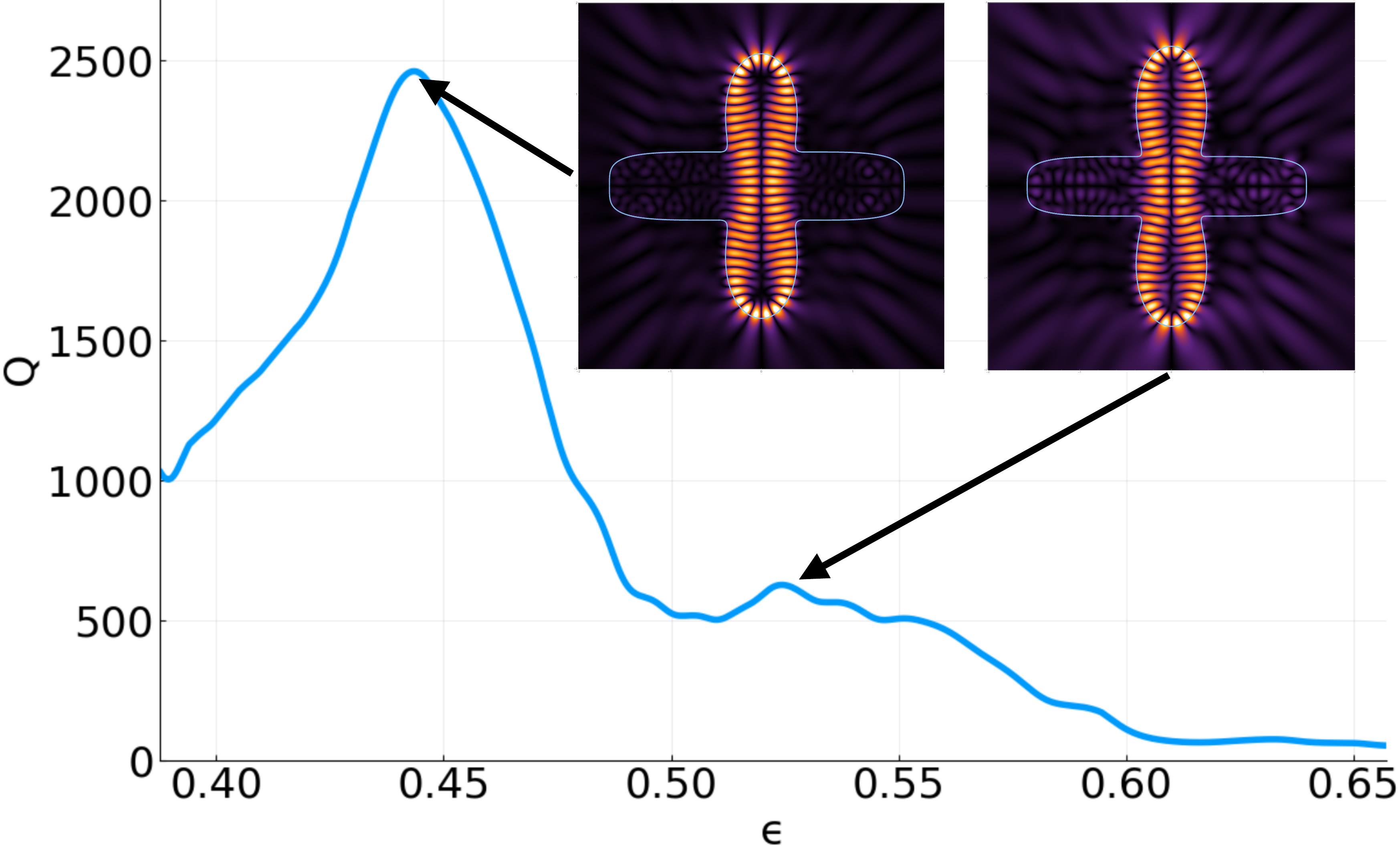}

\caption{\label{fig:Q-factor-versusEps}Spatial structure of a folded chaotic
whispering-gallery mode (insets), and Q factor of the mode versus
deformation parameter $\epsilon$ at refractive index $n=2.4$. The
resonator deformation is defined in Eq.\ (\ref{eq:ShapeCurve}); attached
waveguides of finite lengths are shown horizontal; their lengths are
slightly unequal to remove reflection symmetry. False color represents
field intensity. The real part of $k$ (not shown) decreases approximately
linearly from $14$ to $12.5$ as $\epsilon$ increases.}
\end{figure}

We will describe whispering-gallery type behavior with a different
topology, which can be described as \emph{folded} chaotic whispering-gallery
modes, cf. Fig.\ \ref{fig:Q-factor-versusEps}. This refers to the
fact that the direction of whispering-gallery ray circulation underlying
the modes reverses due to a self-intersection near the center of the
cavity. The bowtie orbit of Ref.\  \citep{Gmachl05061998} also exhibits
a self-intersection, but it -- like all other such orbits we are
aware of -- gives rise to modes that explore the boundary only at
isolated points corresponding to the discrete vertices where ray reflections
occur. In Ref.\  \citep{Gmachl05061998}, $Q>1000$ was obtained at
relatively high refractive index contrast of $n\gtrapprox3$, corresponding
to a critical angle for total internal reflection of $\chi_{\mathrm{c}}\approx\arcsin\nicefrac{1}{n}\approx0.34$
(where $\chi$ denotes the angle of incidence with respect to the surface normal).
For such high index contrast, even Fabry-Perot cavities with near-normal
angle of incidence provide similar Q-factors without the need for
additional mirrors.

By contrast, all the high-Q modes in our design are localized predominantly
in the whispering-gallery region of phase space, much further from
the critical angle, even when the index contrast is below $n=2$.
The light is then well-confined by total internal reflection. Reminiscent
of a figure-eight shape, folded chaotic WGMs have a waist that allows
them to avoid two isolated sections of the boundary; but unlike stable
periodic orbits such as the bowtie, folded chaotic WGMs explore the
remainder of the perimeter in the same way as would be expected for
conventional WGMs. In this sense, these types of modes come as close
as possible to the circulating ray patterns that are strictly ruled
out by Mather's theorem.

The self-intersecting topology makes folded chaotic WGMs amenable
to the incorporation of waveguides because the openings have only
small overlap with the waist of the mode. Although the field looks
similar to Gaussian beam in the waist region, the latter are fundamentally
different because they are always built upon stable periodic orbits,
which in the simplest case requires a configuration corresponding
to two focusing mirrors \citep{BabicBook1972}. In our design, the
reflections are near-grazing along the entire convex part of the boundary,
and there is no focusing-mirror configuration. Therefore, folded chaotic
WGMs cannot be obtained in paraxial optics.

As an important correction to the ray limit in the presence of chaotic dynamics,
{\em dynamical localization} has been invoked \citep{Nockel:1997lr} to explain
the high Q factors of chaotic WGMs in convex cavities, and circular dielectrics
with corrugated surface perturbations \citep{FangCao2005}. A hallmark of
dynamical localization is an exponential decay of wave intensity as a function of
a variable which in the classical ray picture exhibits diffusive time evolution. In
corrugated perturbed circles, this variable is the angular momentum; and it
remains a good quantity to characterize dynamical localization in chaotic WGMs
of smooth but convex cavities \citep{Noeckel:T90-2001}. At each reflection,
the angular momentum is proportional to $\sin\chi$. However, an important
common feature on which previous work relies is that the underlying classical
dynamics exhibits a separation of time scales, without which diffusive behavior
cannot be identified. Before $\sin\chi$ changes significantly, a whispering-gallery
ray will have completed many round trips, allowing the azimuthal positions of
boundary reflections to be averaged out \citep{Noeckel:T90-2001}.

To identify wave localization in the folded WGMs to be discussed here, we
have to follow a different route because angular momentum does not undergo
the required slow diffusion. It jumps discontinuously because a figure-eight
pattern entails periodic reversals of the sense of rotation. Therefore, we begin with
a description of the ray dynamics in phase space before presenting the detailed
results of our wave calculations in section \ref{sec:wavecalc}. These two
descriptions are then synthesized in section \ref{sec:Discussion} to identify
the classical structures on which the modes are localized,
using Husimi projections of the numerical wave functions onto the ray
phase space.

\section{Cavity shape and ray phase space}
\begin{figure}
\includegraphics[width=0.85\columnwidth]{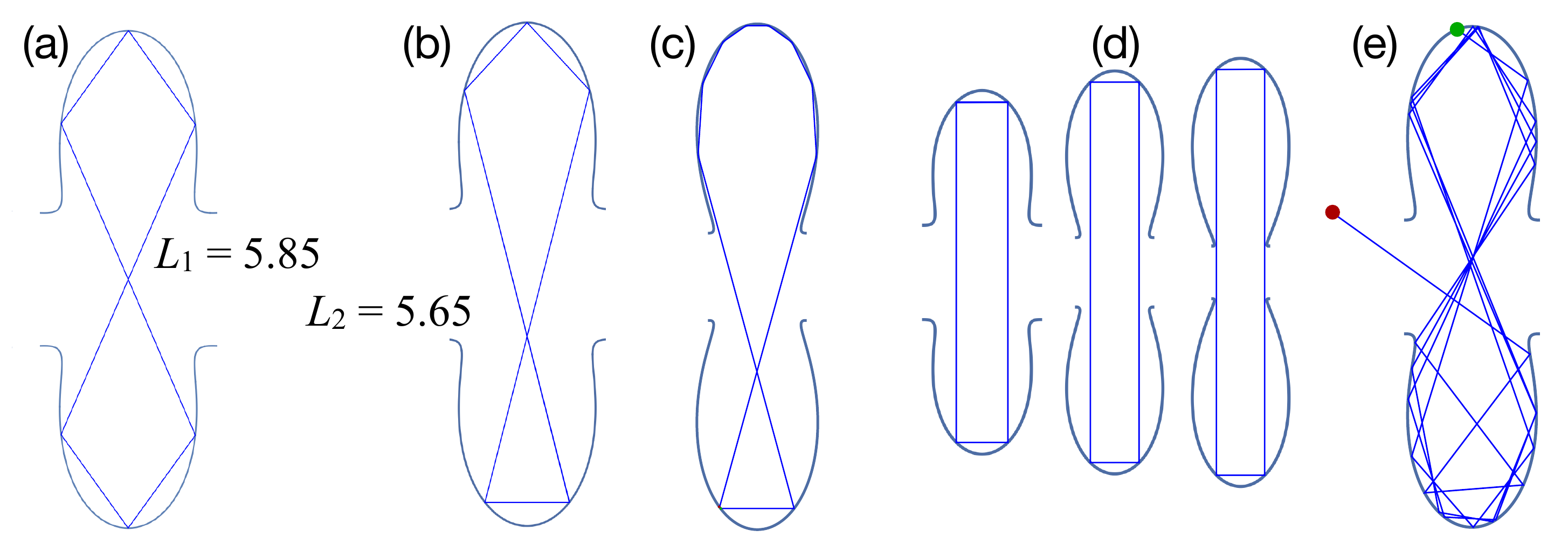}

\caption{(a-c): Three self-intersecting unstable periodic orbits with angles
of incidence $\sin\chi\ge0.6$ in the billiard shape given by Eq.\
(\ref{eq:ShapeCurve}) with $\epsilon=0.43$. The lengths $L_{1}$,
$L_{2}$ of the orbits are slightly different. The same topology of
orbits persists for a wide range of deformations. (c) is obtained
from (b) by adjusting the initial conditions to generate eight instead
of three reflections in the top half. For the rectangular unstable
periodic orbit in (d), the increasing proximity to the corners of
the waveguide aperture is illustrated for increasing $\epsilon$.
(e) A ray trajectory launched in the cavity (green dot) and escaping
through the waveguide opening (red dot). It retains the folded character
of the orbits in (a), (b) for a short time. \label{fig:Two-self-intersecting-periodicOrbits}}
\end{figure}
Figure \ref{fig:Two-self-intersecting-periodicOrbits} illustrates the non-convex
cavity shape in which the long-lived modes are confined. The openings in the
boundary are attached to waveguides, but for the purposes of the ray dynamics they
constitute escape windows, in addition to the refractive escape mechanism that
sets in when the condition for total internal reflection at the dielectric interface
is violated. The three self-intersecting
periodic orbits shown in Fig.\ \ref{fig:Two-self-intersecting-periodicOrbits}
(a - c) have incident angles satisfying $\sin\chi\ge0.6$, which means they are
confined by total internal reflection for refractive indices $n>1.7$.
Half of the reflections occur with opposite sense of circulation, corresponding
to the reversal of angular momentum (with respect to the center)
taking place at the self-intersection.

The geometry also permits a rectangular periodic orbit whose angle
of incidence has the fixed value $\sin\chi\approx0.707$, but Fig.
\ref{fig:Two-self-intersecting-periodicOrbits} (d) shows an important
distinction to the self-intersecting orbits: as $\epsilon$ increases,
the corners of the waveguide aperture encroach on the orbit and eventually
touch it. The resulting corner diffraction \citep{ISI:000323333000016}
will degrade the lifetime of any modes based on this orbit. The numerical
computations to be described in section \ref{sec:wavecalc} have revealed
 high-Q modes in cavities of the shape (\ref{eq:ShapeCurve}), that
do show enhanced intensity near the rectangle orbit, but never exclusively
on that orbit. Instead, the folded chaotic states consistently show
high intensity overlapping with the folded orbits over a wide range
of deformations $\epsilon$ and attached waveguide widths.

\begin{figure}
\includegraphics[width=0.8\columnwidth]{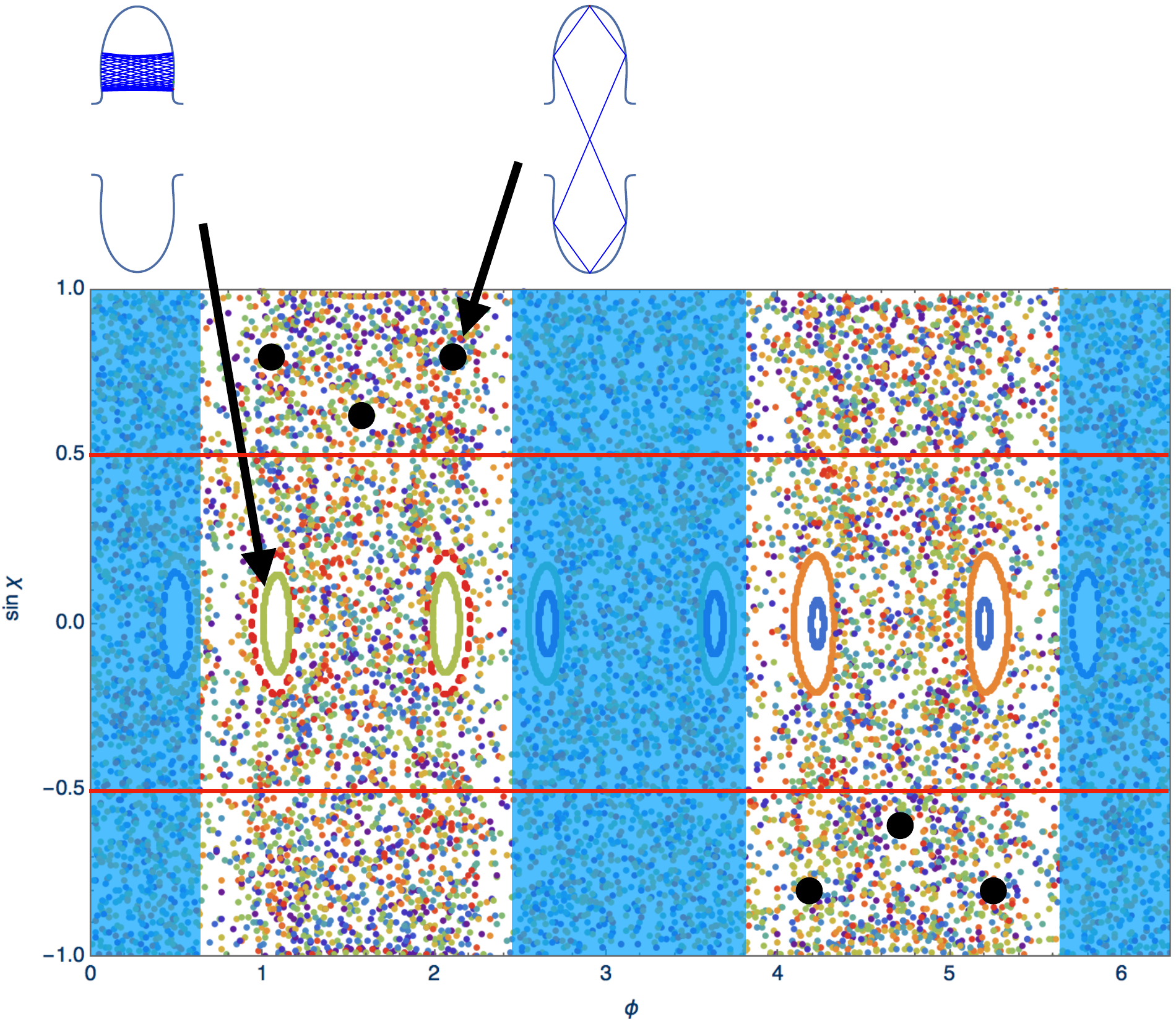}

\caption{\label{fig:Poincare-surface-of-section}Poincar{\'e} surface of section
with the curve parameter $\phi$ as position and $\sin\chi$ as momentum
variable. Here, $\chi$ is the angle between incident rays and the
surface normal, so that $\sin\chi=1$ corresponds to grazing incidence.
Ray trajectories show up as point clouds in the chaotic sea, or as
one-dimensional lines for regular motion. Shown as insets are a stable bouncing-ball
orbit (left) and an unstable periodic orbit (right). Arrows point to the corresponding phase
space locations; in particular the 6 black dots indicate the location of the 6 bounce points
for the periodic orbit.
Shaded boxes indicate the intervals of $\phi$ which describe the attached
waveguides. Rays entering these regions will escape the resonator.
The thin horizontal lines mark the critical angle for total internal
reflection, $\left|\sin\chi_{c}\right|=\nicefrac{1}{n}$ ($n=2$ here). Here and
in all subsequent results, $n$ can be viewed as the interior refractive
index while the exterior refractive index is unity.}
\end{figure}

Figure \ref{fig:Two-self-intersecting-periodicOrbits} (e) illustrates a ray in
the chaotic interior region that escapes
into the waveguide opening after several reflections.
By folding a WGM over itself and attaching waveguides in the avoided
regions of the boundary, two distinct escape windows in phase space
are created: one bounded by the critical angle for refractive escape,
and the other by the real-space locations of the waveguide openings.
This is illustrated in the Poincar{\'e} surface of section of Fig.\ \ref{fig:Poincare-surface-of-section},
depicting the phase space of a billiard parametrized in polar coordinates
by
\begin{equation}
r(\phi)\propto1+\epsilon\,\cos\!\left(4\phi\right)\label{eq:ShapeCurve}
\end{equation}
The curve parameter $\phi$ is recorded on the horizontal axis, and
$\sin\chi$ is a measure of the conjugate momentum variable.

For the numerical wave calculations of Fig.\ \ref{fig:Q-factor-versusEps},
waveguide segments were connected to the open sections of the boundaries depicted
in Fig.\ \ref{fig:Poincare-surface-of-section}. The waveguide segments were modeled as
a ``squircle'' \citep{Squircle}.
This curve has a polar-coordinate representation
\[
r_{\text{WG}}(\phi)=\sqrt{a^{2}\left|\cos\phi\right|+b^{2}\left|\sin\phi\right|}
\]
but was also shifted horizontally in order to connect smoothly with
the unshifted curve of Eq.\ (\ref{eq:ShapeCurve}).

The waveguide portions of the boundary are shaded in the Poincar{\'e}
section because they are irrelevant to the interior ray dynamics that
makes the folded orbits possible.
There are no stable islands associated with any periodic orbits other
than the bouncing-ball trajectories (leftmost inset) whose perpendicular
angle of incidence precludes long-lived stable modes from forming. There are no invariant
curves in the whispering-gallery region near $\sin\chi\to1$.

\section{Wave calculations}\label{sec:wavecalc}

The wave patterns in Fig.\ \ref{fig:Q-factor-versusEps} indicate that
the mode is not supported by any one of the self-intersecting periodic
orbits in Fig.\ \ref{fig:Two-self-intersecting-periodicOrbits} alone, but by a more
extended phase-space region in their vicinity. The main panel of
Fig.\ \ref{fig:Q-factor-versusEps} follows a single mode over a range of
deformations $\epsilon$, revealing an optimal deformation of $\epsilon\approx0.44$
at which this particular mode reaches $Q\approx2500$.

To characterize
the parameter-dependence of the Q factor further, we carry out numerical
simulations of the wave equation using two different methods. The
first approach is a version of the boundary-integral method described
by Heider \citep{HeiderBIM} in which we directly search for the quasibound
states satisfying outgoing-wave boundary conditions at infinity \citep{NockelMcBook1}.
The second approach isolates the high-Q modes by harmonic inversion
 \citep{Mandelshtam:2001fk} of temporal field variation collected
in a finite-difference time domain simulation with perfectly matched
layer boundary condition (MEEP with post-processing by Harminv) \citep{MEEP-citation}.
In the direct quasibound-state calculation, we obtain complex wavenumbers
$k_{\text{QB}}=k-i\,\kappa$ to determine Q using Eq.\ (\ref{eq:QfactorDef}).

\subsection{Quasibound states\label{subsec:Quasibound-states}}

In the boundary-integral approach, we leverage the fact that the dielectric
defining the billiard is uniform so that a Green-function description
of the interior and exterior fields purely in terms of the boundary
is possible. At the dielectric interface, the electric field is assumed
to satisfy the boundary conditions for TM polarization (electric field
perpendicular to the plane). After discretization along the interface,
this leads to a nonlinear eigenvalue problem $A(k_{\text{QB}})\,\boldsymbol{u}=\boldsymbol{0}$
where $\boldsymbol{u}$ contains the source values of the electric
field and its normal derivative, and $A$ is a matrix obtained from
the field matching equations. A non-trivial solution requires searching
for $k_{\text{QB}}$ in the complex plane, which we do using a predictor-corrector
method \citep{HeiderBIM}. The method requires discretization of the
boundary curve, and we generally found good convergence up to wavenumbers
of $k\approx60$ by choosing $580$ points. For additional analysis
of the exponential convergence of the method with discretization density,
see Ref.\  \citep{HeiderBIM}.
\begin{figure}
\includegraphics[width=0.6\columnwidth]{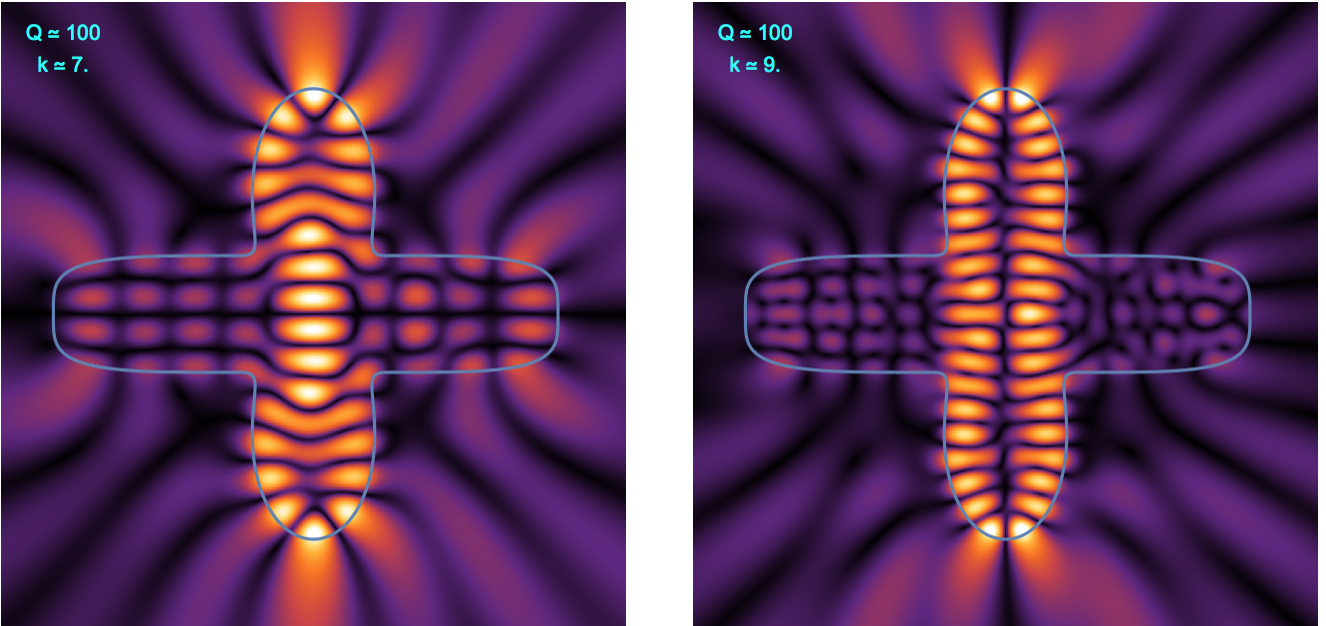}
\vspace*{-0.5cm}
\caption{Small-cavity limit of the folded chaotic WGM, showing only low Q factors
as the transverse width of the mode is comparable to the cavity size.
Below $k\approx7$, the wavelength is too long to observe the WGM
like concentration of intensity near the top and bottom of the cavity.
\label{fig:Small-cavity-limit} The horizontal waveguide stubs are
attached in a way that smoothly matches the shape given by Eq.\ (\ref{eq:ShapeCurve})
for the vertical lobes. The refractive index is $n=2.4$ inside and
$n=1$ outside, and the vertical lobes are described by Eq.\ (\ref{eq:ShapeCurve})
with $\epsilon=0.444$. In this and the following plots, one can
discern even and odd parity with respect to reflections at the horizontal
axis. Although we intentionally break reflection symmetry across the
vertical axis by making the horizontal waveguide lengths unequal,
the modes still show approximate antinodes (left) or nodal lines (right)
along the vertical axis. This is because the cavity supporting most
of the intensity is still left-right symmetric.}
\end{figure}
\begin{figure}
\includegraphics[width=0.75\columnwidth]{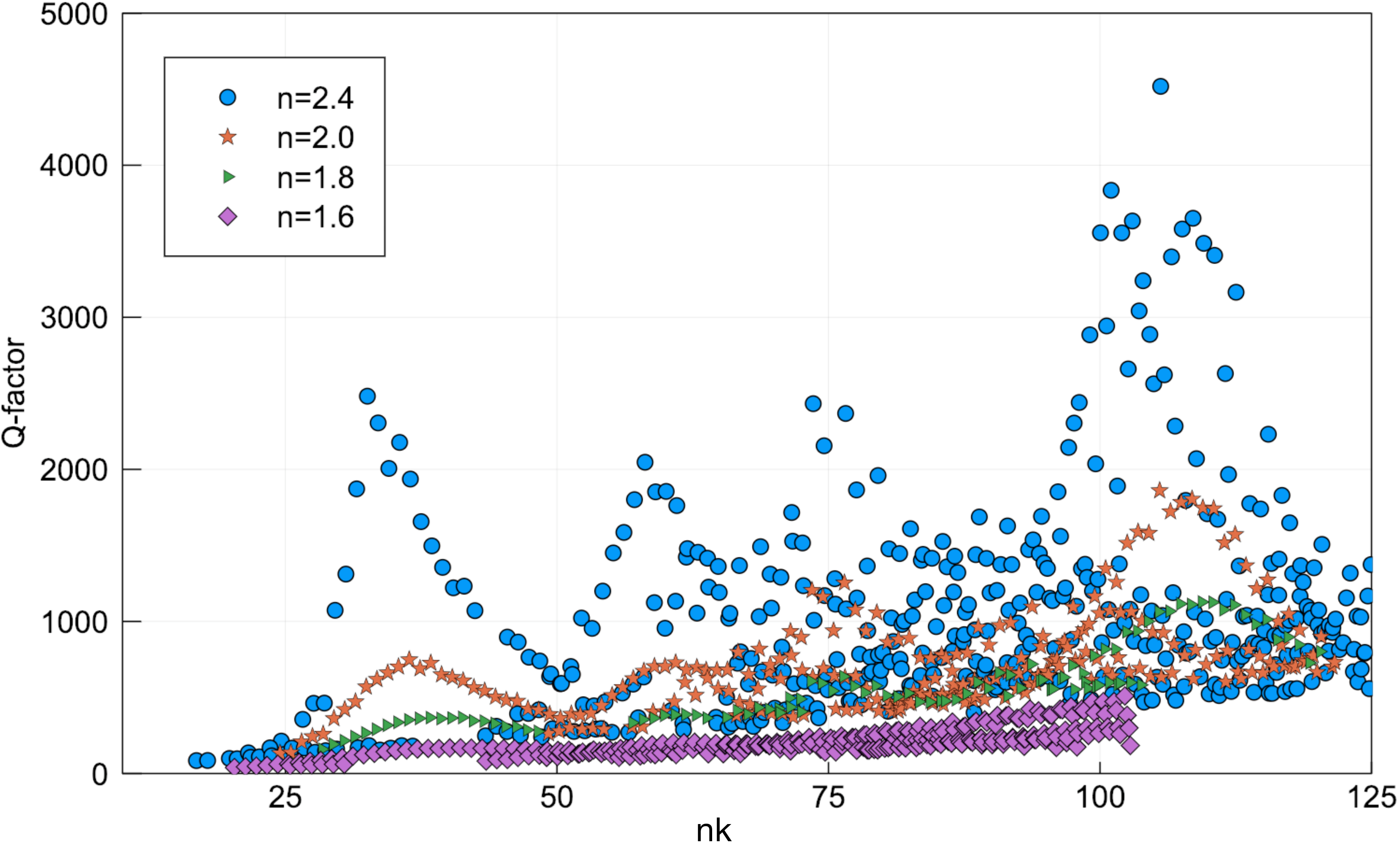}
\vspace*{-0.5cm}
\caption{Wavenumber sweeps of the Q factor for different refractive indices
$n$, discarding low-Q modes with $\kappa\ge0.05$ (for $n=1.6$,
only modes with $\kappa\ge0.2$ are shown). The deformation is $\epsilon=0.444$.
The appearance of distinct Q-factor peaks as a function of quasi-bound
state wavenumber is most pronounced at the largest refractive index,
$n=2.4$, and becomes nearly unobservable at $n=1.6$ (therefore,
data for $n=1.6$ were not collected beyond $nk\approx106.5$). The
peaks that do remain observable are approximately at the same values
of $nk$ for all $n$.\label{fig:Wavenumber-sweeps}}
\end{figure}

By definition, the boundary-integral method assumes that regions of
uniform dielectric constant are bounded by closed curves, so we model
the waveguides as finite-length attachments, see Fig.\ \ref{fig:Q-factor-versusEps}.
By varying the lengths of these waveguides (either together or independently
of each other) we verified that the finiteness of the stubs has no
significant effect on the mode structure: neither real parts nor imaginary
parts of $k_{\text{QB}}$ for the modes with figure-eight topology
were affected unless their Q factor was below approximately $200$
to begin with. Because the modes are built on classical ray orbits,
they will not be found (with appreciable lifetimes) in small cavities.
This is illustrated in Fig.\ \ref{fig:Small-cavity-limit} where the
self-intersecting topology is barely discernible in the wave intensities.
This raises the question whether the Q factors of such modes will
increase \emph{monotonically} with increasing cavity size, or equivalently
with shorter wave length. To address this dependence, we fix the cavity
geometry at $\epsilon=0.444$ to identify all the modes with lifetimes
above a threshold of $\left|\mathrm{Im}(k_{\text{QB}})\right|=\kappa<0.05$
with wavenumber $k\lesssim60$. The particular choice of deformation
corresponds to the maximum in Fig.\ \ref{fig:Q-factor-versusEps} at
$n=2.4$, but the same phenomena are observed at other values of
$\epsilon$.

Figure \ref{fig:Wavenumber-sweeps} shows that the Q factors display
a \emph{non-monotonic} peak structure. The data are plotted for four
different resonators with refractive indices of $n=1.6$, $1.8$,
$2.0$ and $2.4$. The peaks are more pronounced for larger refractive
index because the critical angle for total internal reflection decreases
with $\sin\chi_{\mathrm{c}}=1/n$. By choosing the horizontal axis
to display $nk$ instead of the free-space wavenumber $k$, the peak
positions for different $n$ moreover line up to a good approximation.
Since $nk$ is the wave number inside the cavity, this indicates that
the high-Q peak \emph{locations} are determined by the interior wave
patterns, not the monotonically $n$-dependent coupling to the surrounding
free space. The absolute length scale of the cavity drops out of $Q$
as per Eq.\ (\ref{eq:QfactorDef}).

\begin{figure}
\includegraphics[width=0.8\columnwidth]{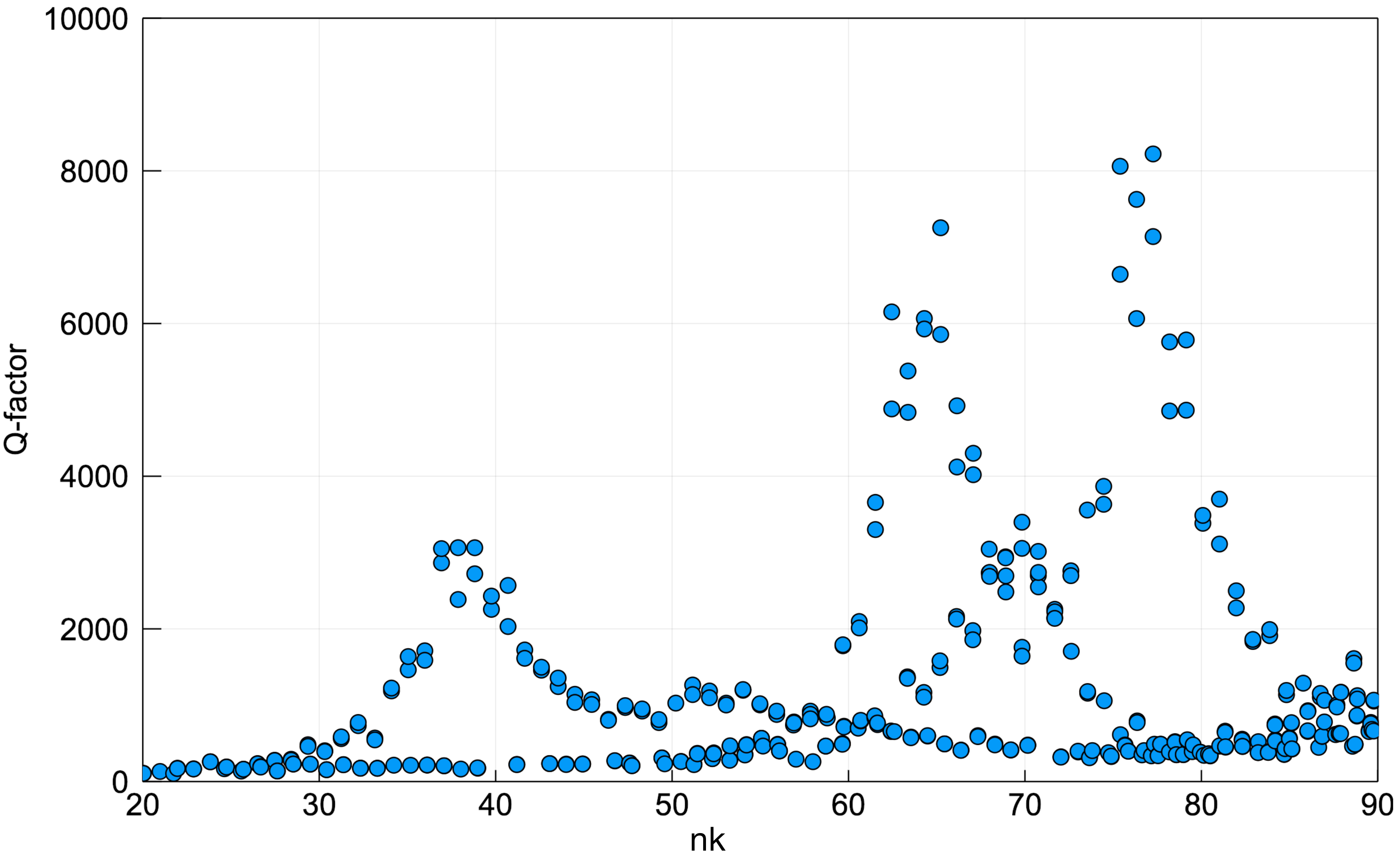}
\vspace*{-0.5cm}
\caption{Q-factor scan versus interior wave number at deformation $\epsilon=0.52$
and refractive index $n=2.4$, showing modes with decay rate $\kappa<0.05$.
\label{fig:Q-factor-scan-e52}}
\end{figure}

The folded chaotic whispering-gallery modes are surprisingly resilient
even to large waveguide openings. For practical applications, much
narrower waveguides will typically be desirable because the input
and output should be single mode. We show results for large openings
because that is the regime in which the coexistence of whispering-gallery
modes with an apparent violation of Mather's theorem is clearest.
In particular, the openings are much wider than the wavelength, so
they cannot be treated as small perturbations. We also observe the
non-monotonic distribution of $Q$-factors shown in Fig.\ \ref{fig:Wavenumber-sweeps}
at $\epsilon=0.444$ for other deformations and waveguide widths.
Figure \ref{fig:Q-factor-scan-e52} shows this for a higher deformation
of $\epsilon=0.52$.

\begin{figure}
\includegraphics[width=0.6\columnwidth]{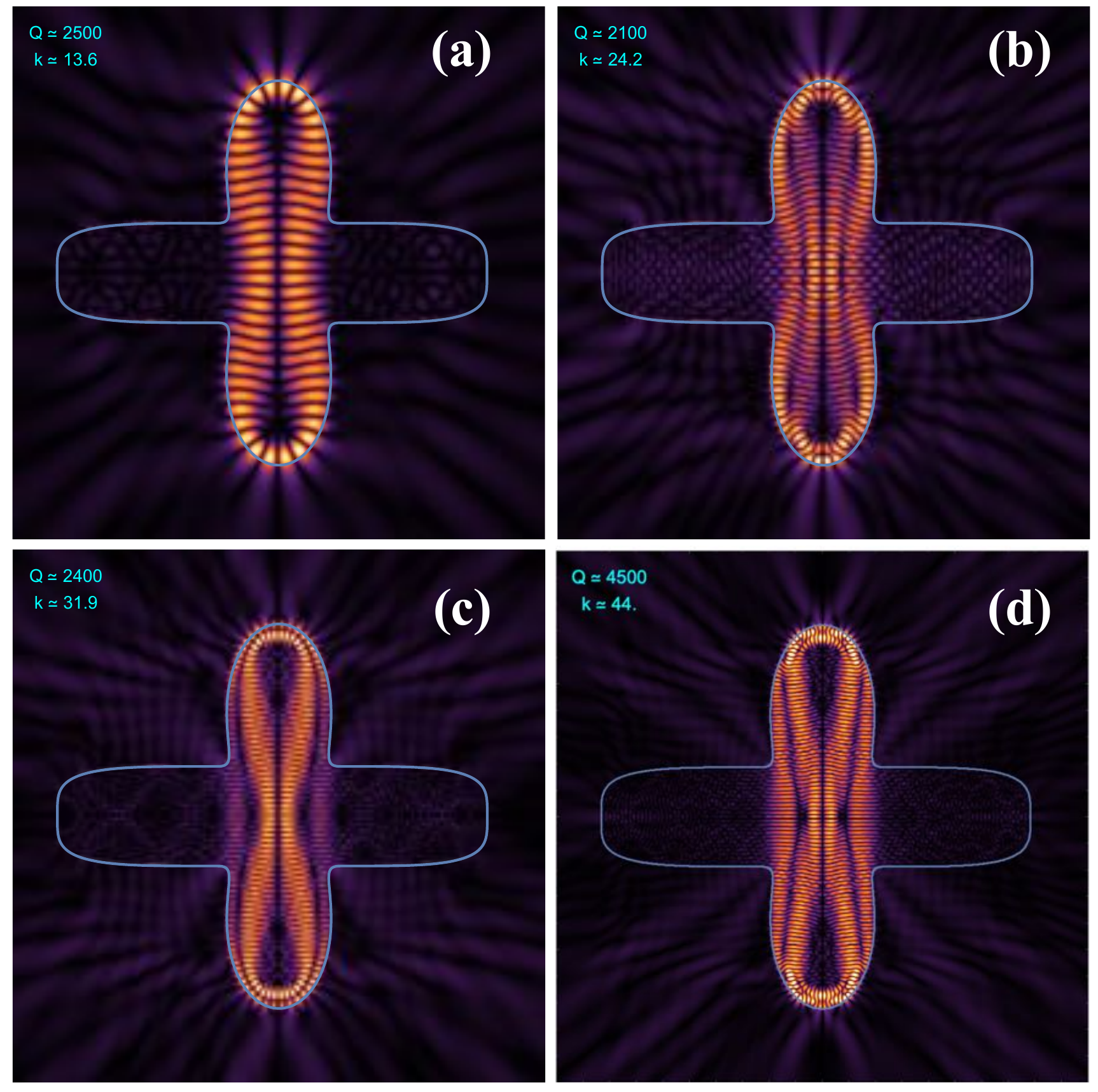}
\vspace*{-0.5cm}
\caption{Field intensities for the modes corresponding to the first four Q-factor
peaks in Fig.\ \ref{fig:Wavenumber-sweeps} for $n=2.4$, $\epsilon=0.444$. The Q factors are
(a) $2500$, (b) $2100$, (c) $2400$ and (d) $4500$.
\label{fig:Field-intensities-forPeaks}}
\end{figure}

Figure \ref{fig:Field-intensities-forPeaks} depicts the wave intensities
of the highest-Q folded chaotic WGMs in Fig.\ \ref{fig:Wavenumber-sweeps},
with wavenumbers ranging from $k\approx13.6\ldots44$ at $n=2.4$.
Each image represents a single quasibound state, not a superposition.
The nodal structure that decorates the underlying figure-eight patterns
becomes more complex with increasing $k$, in much the same way that
transverse nodes appear in higher-order Gaussian beams or radial nodes
appear in whispering-gallery modes. The difference between those examples
and the folded chaotic WGMs is that the nodal lines show wave dislocations
typical of non-separable wave equations. Non-separability goes along
with the chaotic ray dynamics of Fig.\ \ref{fig:Poincare-surface-of-section}.
Despite the increasingly complex nodal structure, all modes share
the whispering-gallery like wave propagation along the convex parts
of the boundary.

\begin{figure}
\includegraphics[width=0.9\columnwidth]{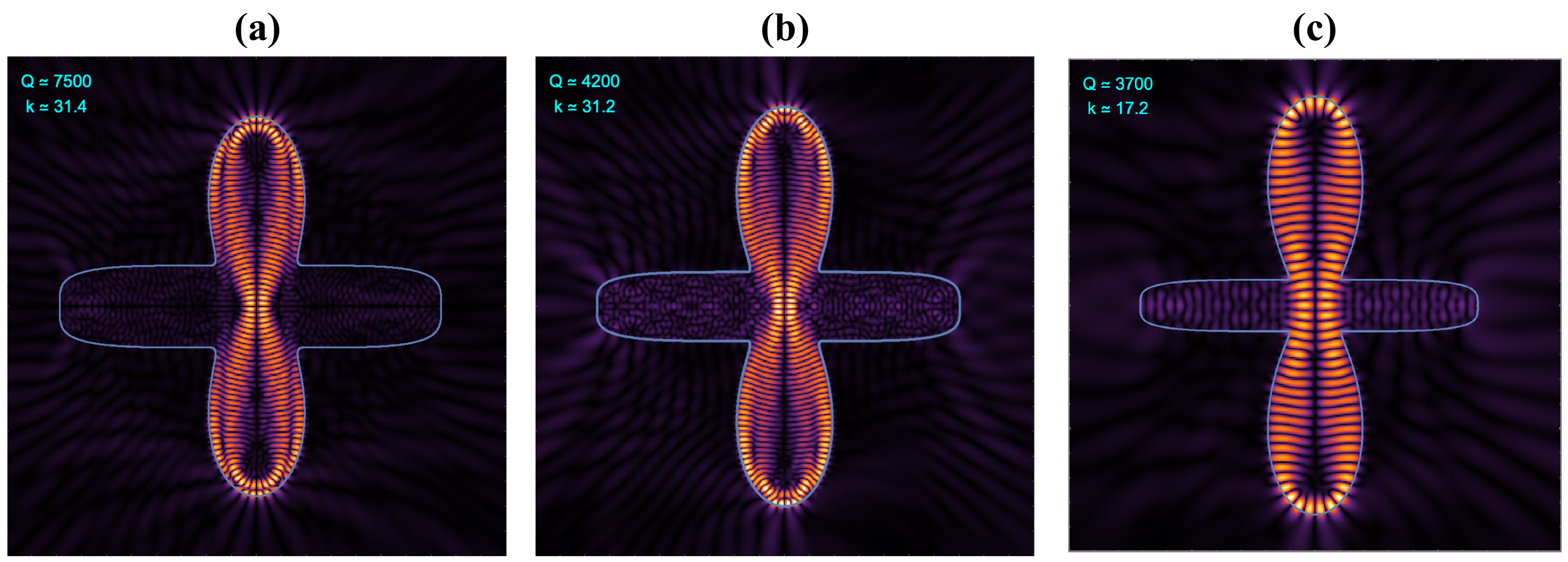}
\vspace*{-0.5cm}
\caption{Field intensities for folded WGMs at different deformations: (a) $\epsilon=0.52$,
(b) $\epsilon=0.6$, (c) $\epsilon=0.7$. The refractive index
is $n=2.4$. In (c), a smaller wavenumber is chosen, whereas (a) and
(b) have comparable wavenumbers. The narrowing of the horizontal stubs
is a result of the requirement that its tangents must match the curve
described by Eq.\ (\ref{eq:ShapeCurve}) at the corners. \label{fig:Field-intensitiesVaryEps}}
\end{figure}

The Q-factor of any given mode may depend non-monotonically on the
deformation parameter $\epsilon$ of Eq.\ (\ref{eq:ShapeCurve}), as
shown in Fig.\ \ref{fig:Q-factor-versusEps}. However, we find other
high-Q modes over the entire range of deformations in Fig.\ \ref{fig:Q-factor-versusEps},
as illustrated by the examples in Fig.\ \ref{fig:Field-intensitiesVaryEps}.
The robustness of the folded morphology against deformation is another
interesting feature that these modes share with conventional whispering-gallery
modes in convex resonators.

\subsection{Time-domain results}

\begin{figure}
\includegraphics[width=0.8\columnwidth]{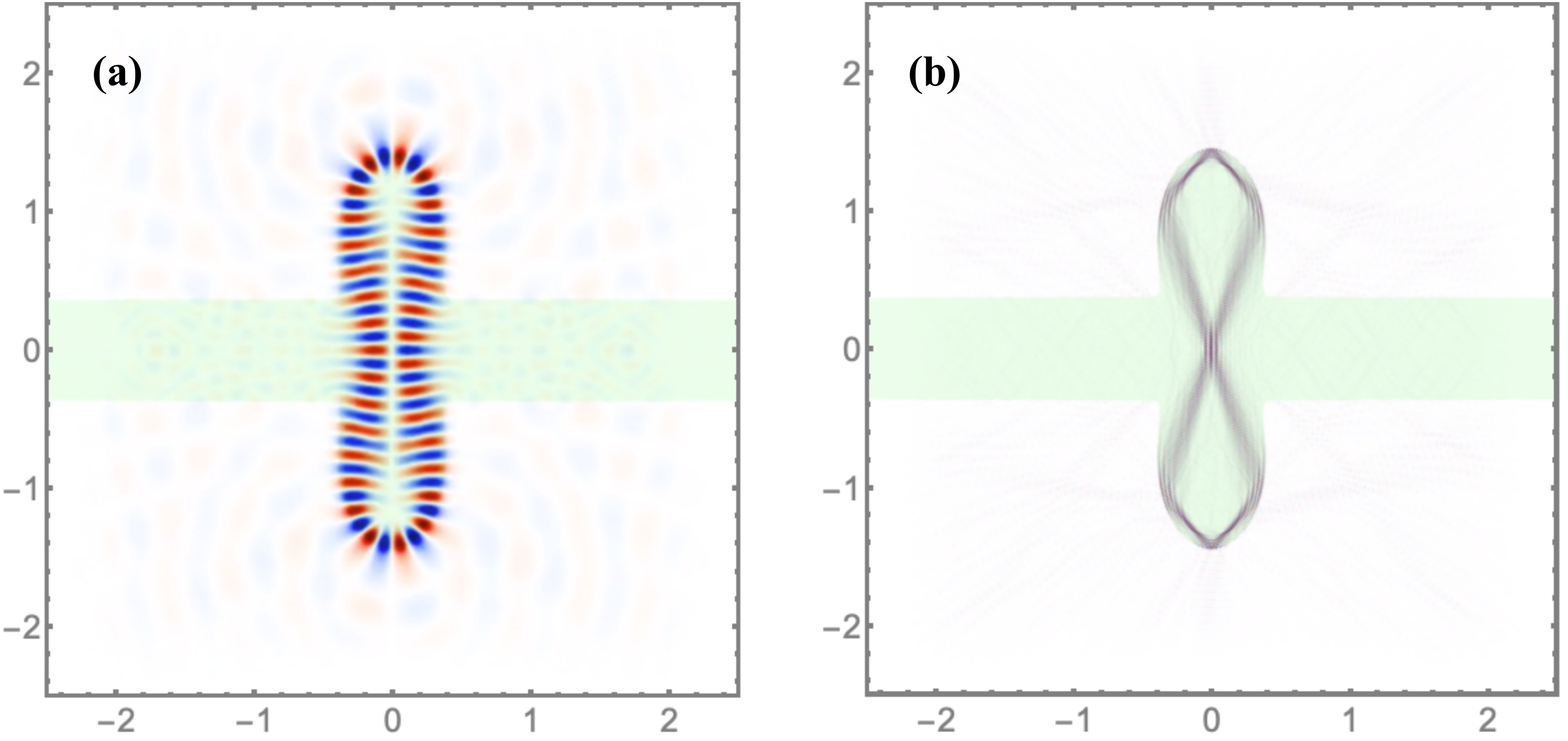}
\vspace*{-0.5cm}
\caption{(a) Folded chaotic whispering-gallery mode as obtained in a finite-difference
time domain computation for the same deformation $\epsilon=0.444$
and refractive index $n=2.4$ shown in Fig.\ \ref{fig:Wavenumber-sweeps}.
The wave number corresponds to the top of the first Q-factor peak,
$k\approx13.6$ ($Q=2612$). Waveguide and resonator structure are
underlaid as green shading, and the false-color scale represents the
electric field. (b) Same deformation and refractive index, but approximately
ten times shorter wavelength, $k\approx116.0$ and $Q=9898$. (b)
is discussed in section \ref{sec:Discussion}.\label{fig:MEEPmode1}}
\end{figure}

To further validate the results of the boundary-integral method, we
repeated the calculations for some of the modes using MEEP, which
in particular allows for a different modeling of the waveguide attachments.
Instead of assuming them to be finite stubs, we allowed them to extend
to the boundary of the simulation domain which includes a perfectly
matched layer that prevents back-reflections of a wavepacket which
was launched in one arm of the structure. After settling into a late-time
decaying state, the wave field of a typical high-Q mode displays the
same pattern found in the previous section, cf. Fig.\ \ref{fig:MEEPmode1}
(a). Shown in Fig.\ \ref{fig:MEEPmode1} (b) is the same resonator
at a roughly tenfold shorter wavelength. The comparison illustrates
why we make a distinction between folded chaotic WGMs as in Fig.\ \ref{fig:MEEPmode1}
(a) and scarred states. The concentration of intensity on the six-bounce
unstable periodic orbit of Fig.\ \ref{fig:Two-self-intersecting-periodicOrbits}
(a) identifies the mode of Fig.\ \ref{fig:MEEPmode1} (b) as a scarred
state, whereas there is no single periodic orbit that describes the
intensity of a folded chaotic WGM.

\begin{figure}
\includegraphics[width=0.8\columnwidth]{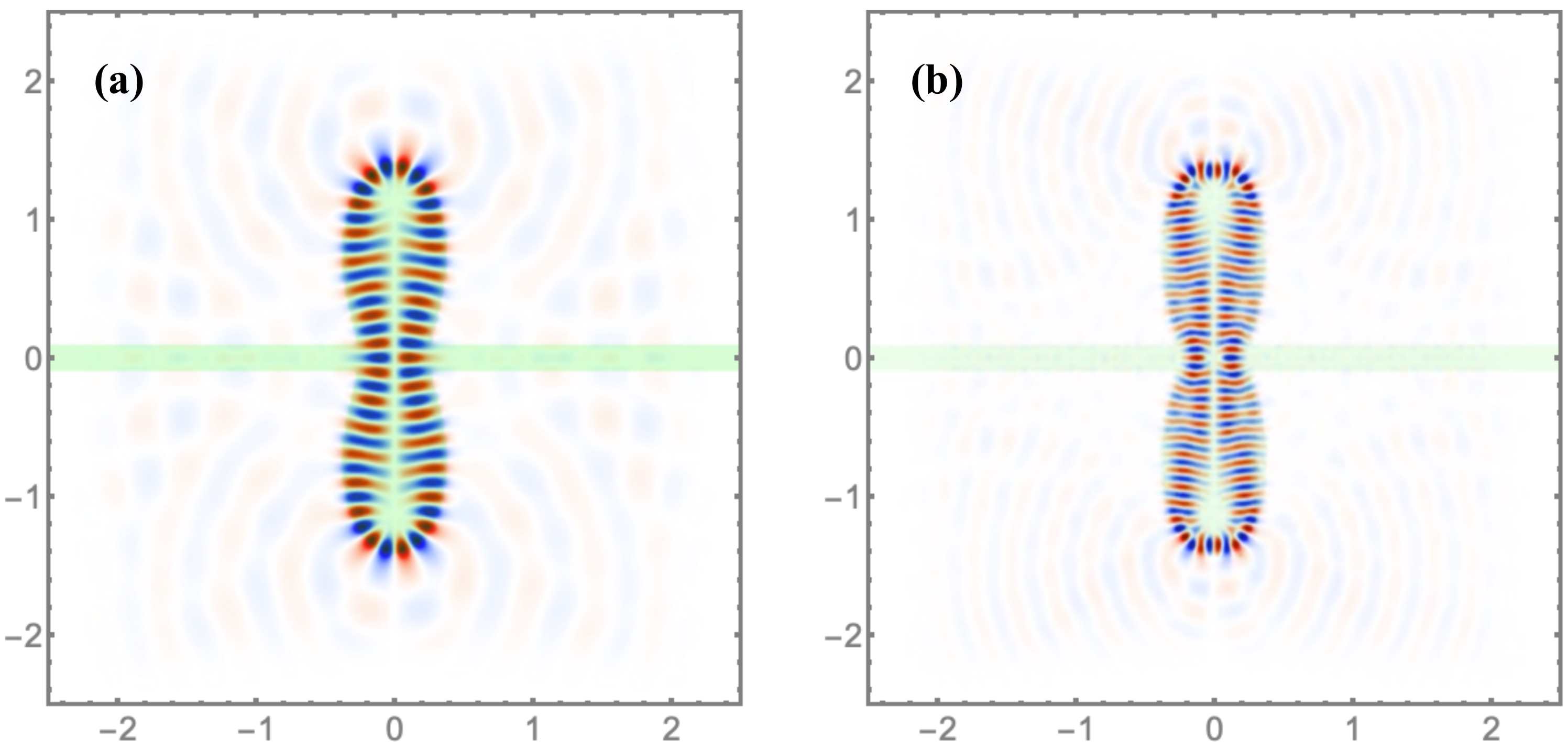}
\vspace*{-0.5cm}
\caption{Two different modes with thinner waveguides, obtained by introducing
a vertical offset between the two lobes of Eq.\ (\ref{eq:ShapeCurve})
and adjusting$\epsilon$ to match them smoothly to the horizontal
waveguides. The refractive index is $n=2.4$. In (a) $k\approx13.3$
and $Q\approx1580$, in (b) $k\approx22.4$ and $Q\approx2550$. \label{fig:Two-different-modesThinWires}}
\end{figure}

Having convinced ourselves that the same results can be obtained with
both numerical approaches (boundary-integral and finite-difference
time domain), we proceed with MEEP to look for folded chaotic WGMs
in resonators with thinner waveguides, cf. Fig.\ \ref{fig:Two-different-modesThinWires}.
The size of the opening in Fig.\ \ref{fig:Two-different-modesThinWires}
(a) is comparable to the wavelength, whereas it is approximately twice
the wavelength in (b). The whispering-gallery circulation along the
boundary, characteristic of the earlier results in Fig.\ \ref{fig:Q-factor-versusEps},
shows that the phenomenon is robust not only to variations in the
refractive index, but also to changes in waveguide width.

\section{Discussion\label{sec:Discussion}}

\subsection{Folded chaotic whispering-gallery modes in phase space}

Individual modes of the folded chaotic WGM type do not shift or broaden
significantly when the attached waveguides are shortened into stubs.
From this we conclude that leakage into the waveguides is not the
limiting factor for the lifetimes of individual folded chaotic WGMs.
However, this is not straightforward to reconcile with the ray-tracing
results depicted in Figs.\ \ref{fig:Poincare-surface-of-section} and
\ref{fig:Two-self-intersecting-periodicOrbits} (e):

A comparison between the numerically observed peak Q factors in Fig.
\ref{fig:Wavenumber-sweeps} and the classical ray escape into the
waveguides provides a first indication that dynamical wave localization
is essential in the formation of folded chaotic WGMs. We find that
up to a set of measure zero (the trapped unstable periodic orbits),
no matter where in the classical phase space we launch a ray, it escapes
into the waveguide openings much too fast to explain the $Q$ factor
found in the wave calculations.

The rapid ray escape can be viewed as a result of Mather's theorem,
because the non-convex billiard shape implies that the bounce dynamics
does not constitute a twist map, and consequently neither the Lazutkin
nor the Poincar{\'e}-Birkhoff theorems apply \citep{LazutkinBook}. What
remains is a largely chaotic phase space as in Fig.\ \ref{fig:Poincare-surface-of-section},
and no stable ray orbits around which high-Q modes can be form by
the mechanism of paraxial optics \citep{BabicBook1972}.

\begin{figure}
\includegraphics[width=\columnwidth]{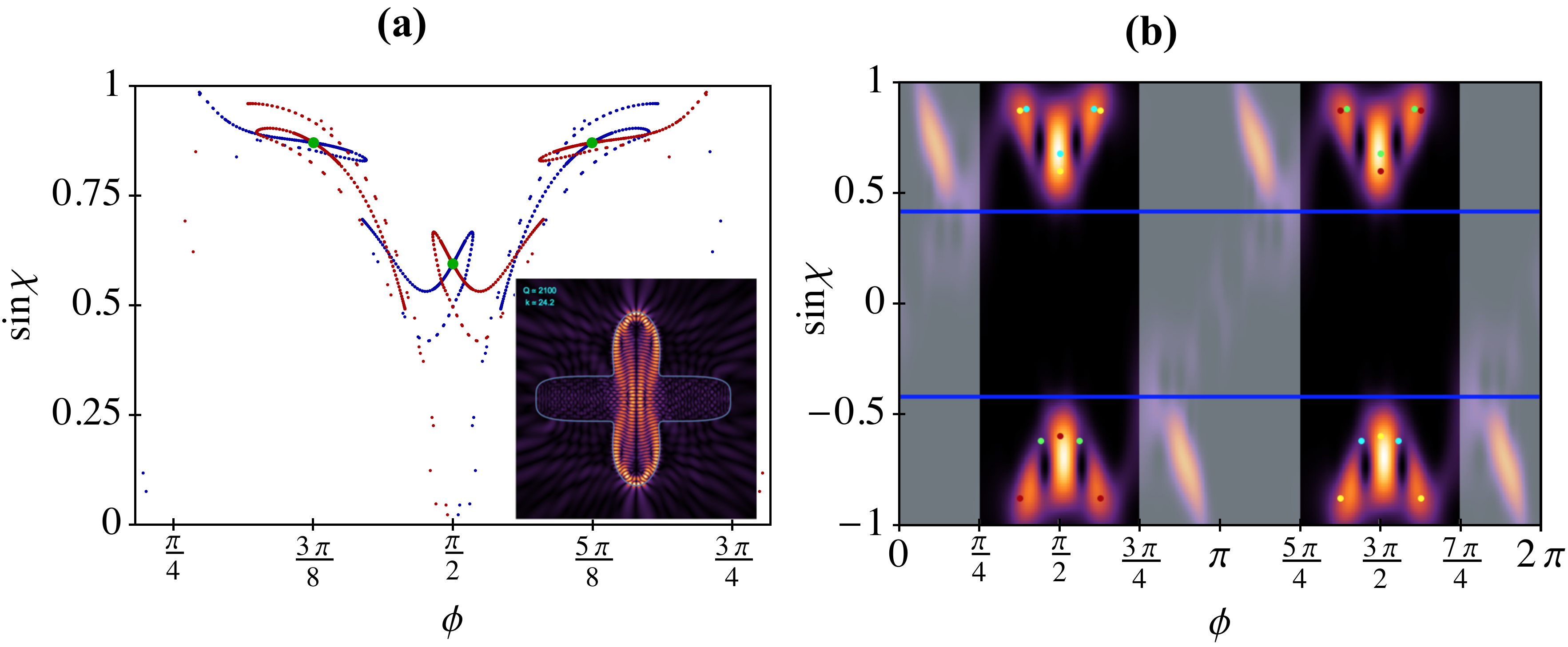}
\vspace*{-0.5cm}
\caption{(a) Stable and unstable manifolds around an unstable periodic orbit.
The zoomed-in region shown in (a) corresponds to the top left quadrant
of the Husimi projection in (b). The real-space wave intensity of
the mode in (b) is shown in the inset to (a). It is the same mode as shown in Fig.\ \ref{fig:Field-intensities-forPeaks} (b). Shaded
regions in (b) mark the waveguides, which were treated as escape windows
in (a). The thin horizontal lines mark $\left|\sin\chi_{c}\right|=\nicefrac{1}{2.4}$. Refractive escape is not considered in the ray simulation
for (a), to get a more complete picture of the manifolds. Note the
correspondence in shape between the phase space structure and the
areas of large Husimi weight.\label{fig:HusimiPhaseSpace}}
\end{figure}

However, as suggested by Fig.\ \ref{fig:Two-self-intersecting-periodicOrbits}
(e), even in a chaotic region of phase space, the motion is nevertheless
organized by the periodic orbits: Each unstable periodic orbit is
a periodic point of the Poincar{\'e} map which has stable and one unstable
manifolds. In the linear regime near a periodic point, they correspond
to trajectories that either converge on, or recede from, that point.
By launching a large number of rays from a small neighborhood of the
periodic point and iterating the billiard map forward and backward
in time, these manifolds are traced out, revealing a characteristic
web of intersections \citep{LichtenbergBook1992} -- the homoclinic
tangle. In the main panel of Fig.\ \ref{fig:HusimiPhaseSpace} (a),
this has been done for the six-bounce orbit corresponding to the thick
solid dots in Fig.\ \ref{fig:Poincare-surface-of-section}. Embedded
in the intersections between the manifolds are higher-order periodic
points corresponding to orbits such as the one shown in Fig.\ \ref{fig:Two-self-intersecting-periodicOrbits}
(c). Guided by the homoclinic tangle, the phase space flow is far
from random on intermediate time scales and in fact helps explain
the directional emission from free-standing asymmetric cavities \citep{NoeckelChapter2002}
by predicting at what positions the condition for total internal reflection
is first violated for a chaotically diffusing trajectory.

To expose the relationship between the observed modes and the underlying
ray phase space structure exemplified in Fig.\ \ref{fig:Poincare-surface-of-section},
it is useful to project the numerical wave results onto the phase
space by means of the Husimi function. Choosing $\phi$ and $\sin\chi$
as the coordinates for this projection, only the boundary fields are
needed, and this is just what the boundary-integral approach provides
 \citep{PhysRevE.90.022903}. Therefore, Fig.\ \ref{fig:HusimiPhaseSpace}
shows one of the high-Q modes as obtained with this method.
In Fig.\ \ref{fig:Wavenumber-sweeps} it corresponds to the top of the peak structure
located at $n k \approx 58$ as well as the mode depicted in Fig.\ \ref{fig:Field-intensities-forPeaks} (b).
The wave intensity shown in the inset illustrates the appearance of
nodal lines which in a conventional WGM would be the radial zeros.
As seen in Fig.\ \ref{fig:Field-intensities-forPeaks}, successive
Q-factor peaks at higher $nk$ in Fig.\ \ref{fig:Wavenumber-sweeps}
show additional such ``radial'' nodes, but the chaotic nature of
the underlying phase space makes a rigorous classification in terms
of radial nodal lines ambiguous.

The Husimi plot in Fig.\ \ref{fig:HusimiPhaseSpace} (b) instead classifies
the mode according to the region of phase space by which it is supported
-- a procedure that is especially useful in this case because there
are no stable ray orbits in the regions of the Poincar{\'e} section bounded
by the escape conditions. The six-bounce periodic orbit whose manifolds
are explored in Fig.\ \ref{fig:HusimiPhaseSpace} (b) is also shown
in Fig.\ \ref{fig:HusimiPhaseSpace} (b), colored yellow and red overlaying
the Husimi intensity of the wave solution (the two colors belonging
to the two opposite senses in which the figure-eight is traversed).
Shown in green and cyan are the self-intersecting period-five orbits.
We have identified other periodic orbits with four to six bounces
that are confined by total internal reflection in the same area of
phase space, and it is not possible to uniquely assign regions of
high Husimi intensity to a single orbit. However, Fig.\ \ref{fig:HusimiPhaseSpace}
(b) does indicate unambiguously that the wave solution is in fact
extended over a region of phase space bounded away from the critical
angle $\chi_{c}$ by the V-shaped tangle of manifolds of Fig.\ \ref{fig:HusimiPhaseSpace}
(a).

\begin{figure}
\includegraphics[width=0.5\columnwidth]{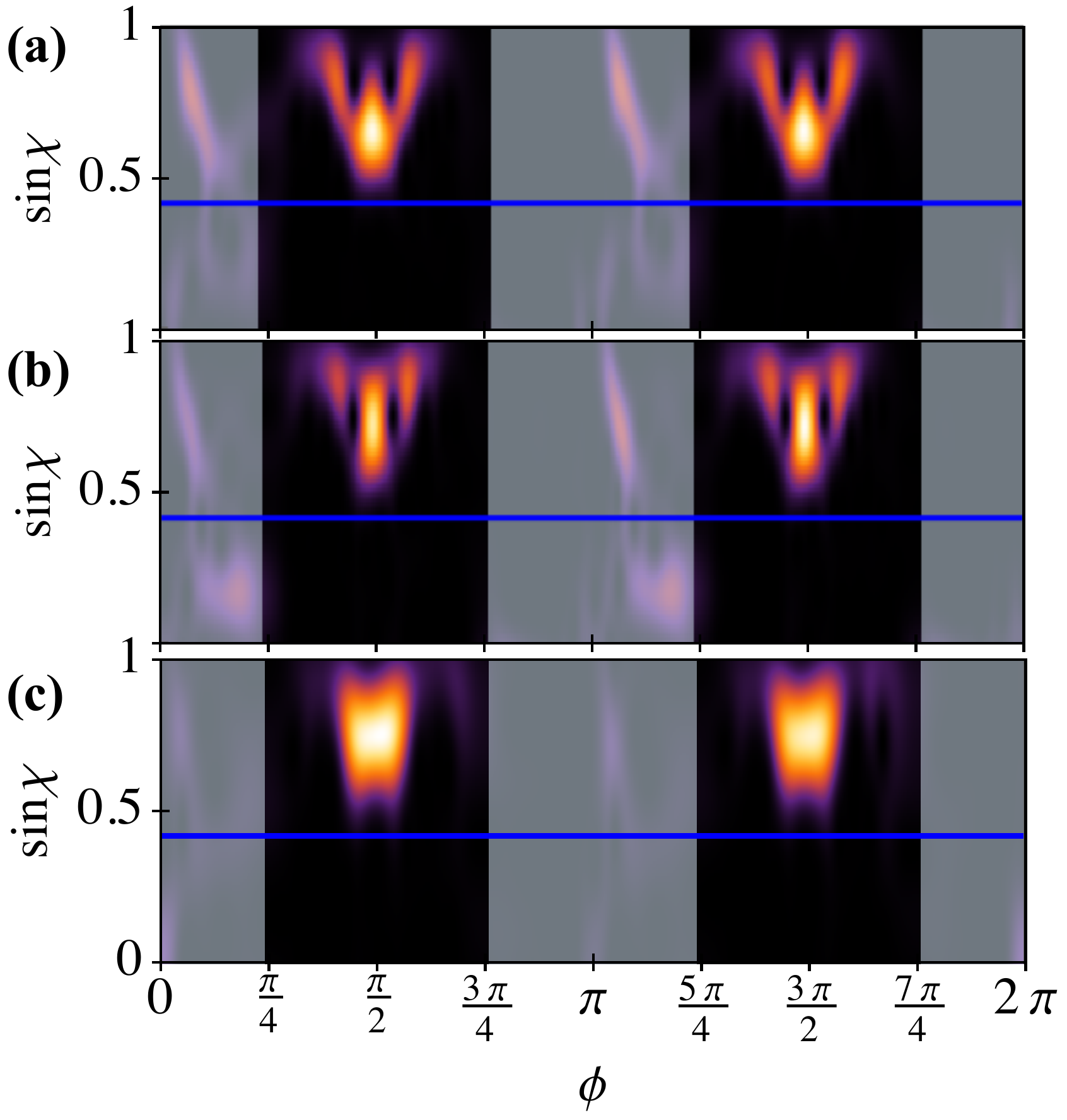}
\vspace*{-0.5cm}
\caption{Comparison of Husimi plots for the same states shown in Fig.\ \ref{fig:Field-intensitiesVaryEps}.
Only the positive-$\sin\chi$ half of the phase space is displayed (shaded regions and horizontal lines
as in Fig.\ \ref{fig:HusimiPhaseSpace}).
The low wavenumber ($k\approx17.2$) in (c) leads to lower phase-space
resolution than in (a) and (b) where $k\approx31.4$ and $k\approx31.2$.\label{fig:Comparison-of-Husimi}}
\end{figure}

Figure \ref{fig:Comparison-of-Husimi} illustrates the common phase-space
region shared between all high-Q modes. Just like the real-space wave
functions, the Husimi intensity shows some variability between different
modes. In particular, the areas of highest Husimi intensity are not centered
on a single orbit of the many figure-eight type periodic orbits that exist in the cavity.

When interpreting the Husimi projection, the question arises whether
our inability to assign a given mode to a unique unstable periodic
orbit is merely a consequence of phase-space resolution. If this were
true, it would mean the all the high-Q modes in our geometry are really
of the scarred kind shown in Fig.\ \ref{fig:MEEPmode1} (b). However,
this is not the conclusion we draw. With the length scale used here,
$k$ ranges from $k\approx7$ in Fig.\ \ref{fig:Small-cavity-limit}
to $k\approx116$ in Fig.\ \ref{fig:MEEPmode1} (b). Within this approximate
window, scarred modes belonging to a single unstable periodic orbit
were not found below $k\approx70$. The common feature of those lower-$k$
modes is instead the appearance of a caustic-like concentration of
intensity following the curvature of the surface at $\phi\approx\pi/2,\,3\pi/2$
(the top and bottom in the wave plots).

In a convex billiard, whispering-gallery circulation leads to true
caustics, but there is also a dense set of periodic orbits in the
shape of inscribed polygons with lengths that accumulate at the circumference
of the billiard. This particular property of inscribed polygonal orbits
in the whispering-gallery region of phase space finds its counterpart
in our non-convex geometry: there is a similar accumulation of periodic
orbits with figure-eight topology, characterized by an increasing
number of bounces along \emph{either} the top \emph{or} the bottom
portion of the billiard. The orbit displayed in Fig.\ \ref{fig:Two-self-intersecting-periodicOrbits}
(c) is an example.

Two other possible scenarios are known to produce wave solutions localized
on phase-space regions where the classical billiard shows no stable
structure: marginally stable periodic orbits \citep{WiersigAnticrossingModes,CaoWiersigReview},
and ray orbits that become stabilized only due to the ``softness''
of the dielectric billiard boundary as a function of incident angle
 \citep{Foster:07,Unterinninghofen2008}. In these cases, the wave
intensity is ``scar-like'' in the sense that it again coalesces
onto identifiable periodic orbits, which is not the case for the folded
chaotic WGMs.

Given that purely ray based escape rates as shown in Fig.\ \ref{fig:Two-self-intersecting-periodicOrbits} (e) are
too fast to explain the high Q factors, we therefore conclude that the folded chaotic WGMs are dynamically
localized on the homoclinic tangle of the self-intersecting periodic orbits in this non-convex cavity.

\subsection{Q-factor oscillations\label{subsec:Q-factor-oscillations}}

The results shown in Figs.\ \ref{fig:Wavenumber-sweeps} and \ref{fig:Q-factor-scan-e52}
are of particular practical importance: in designing resonators that
exhibit modes with a folded topology, the geometric shape alone doesn't
guarantee that $Q$ factors will increase monotonically with $k$,
so the operating wavelength is constrained to match one of the $nk$
ranges where modes with high $Q$ can be found. Whispering-gallery
modes of the perfect circle do \emph{not} show any comparable grouping
into peaks of higher and lower $Q$; instead, maximum attainable $Q$
factors grow monotonically with $nk$ because the modes with the lowest
radial node number grow more and more concentrated at the surface,
corresponding to grazing angle of incidence. Therefore, Figs.\ \ref{fig:Wavenumber-sweeps}
and \ref{fig:Q-factor-scan-e52} show an aspect of the folded modes
that distinguishes them from the conventional WGMs in convex resonators.

By varying the widths and lengths of the attached waveguides, we
ruled out that the periodicity in $nk$ is due to threshold effects
associated with the opening of new propagation channels. In an attempt
to understand the $Q$-factor oscillations at the level of individual
modes, we have singled out some high-$Q$ examples for closer investigation.
In Fig.\ \ref{fig:Wavenumber-sweeps}, the highest-$Q$ mode occurs
at $nk\approx105.6$. It is also shown in Fig.
\ref{fig:Field-intensities-forPeaks} (d) at $k\approx44$. By following
this mode over a neighborhood of deformation parameters $\epsilon$
around the shown values, we found that no crossings or anticrossings
with other modes occur. Such anticrossings can in principle lead to
non-monotonic $Q$-factors as a function of a system parameter, in
particular when whispering-gallery and chaotic modes coexist because
they often depend differently on said parameters \citep{Hackenbroich97}.

Husimi plots provide phase-space information about the classical structures
supporting the folded chaotic WGMs, but their resolution is limited
by the Fourier uncertainty relation between position $\phi$ and angular
momentum as measured by $\sin\chi$ \citep{NoeckelChapter2002}. Individual
modes also vary in the relative weighting between the neighborhoods
of the five- and six-bounce orbits, suggesting that an analysis of
the Q-factor oscillations in Fig.\ \ref{fig:Wavenumber-sweeps} is
best done not at the level of individual modes, but by adopting a
more global point of view.

Such a global perspective can often be obtained by investigating the
connection between the classical periodic orbits and the density of
states \citep{BrackBook1997}. In open systems, the density of states
is continuous but can exhibit resonant structure because it is proportional
to the Wigner-Smith delay time \citep{NoeckelChapter2002}. Oscillations
in the number of high-$Q$ modes could be interpreted as oscillations
in a density of states in which short-lived resonances are not counted.
In order to relate these oscillations to periodic orbits, we explored
the approach of reference \citep{PhysRevLett.70.568}, where beats
between a small number of trajectories with similar actions in the
periodic orbit sum were successfully used to interpret structure in
the density of states for a H{\'e}non-Heiles Hamiltonian. Using the period-orbit
lengths $L_{1}$ and $L_{2}$ from Fig.\ \ref{fig:Two-self-intersecting-periodicOrbits}
(a, b) to obtain the actions $S_{1,2}=nkL_{1,2}$, we arrive at an
estimate for the beating period in Fig.\ \ref{fig:Wavenumber-sweeps}
of
\begin{equation}
\Delta(nk)\approx\frac{2\pi}{\left|L_{1}-L_{2}\right|}\approx30\label{eq:DeltaKnperiod}
\end{equation}
at refractive index $n=2.4$. Although this is in reasonable agreement
with the spacing between the first two peaks in Fig.\ \ref{fig:Wavenumber-sweeps},
we have not yet been able to reproduce the correct dependence of the
numerically observed peak positions on deformation parameter $\epsilon$.
In particular, we have found additional periodic orbits with larger
values of $\left|L_{1}-L_{2}\right|$ that should appear with comparable
amplitudes in the periodic-orbit sum for the density of states, so
it remains to be seen if a refined version of Eq.\ (\ref{eq:DeltaKnperiod})
will be able to preserve the appealing simplicity of this interpretation
in terms of periodic-orbit beats.

\section{Conclusions}

We have numerically calculated the quasi-bound state wave functions
of a series of planar dielectric cavities with attached waveguides
that support the formation of folded chaotic whispering-gallery modes.
We considered a large range of wavelengths and refractive indices,
using shape deformations that combine two convex halves in a non-focussing
configuration. It is possible to sustain useful quality factors ($10^{3}$
or larger) despite the fact that the ray dynamics permits long-lived,
trapped trajectories only for a discrete set of unstable periodic
orbits. The figure-eight topology of these orbits imprints itself
on the wave solutions even though there is no one-to-one correspondence
between individual modes and single periodic orbits.

As a result of this folding, the modes are confined by total internal
reflection in the uninterrupted convex parts of the resonator boundary,
while at the same time exhibiting a waist that reduces their overlap
with the opening to the attached waveguides. Because these openings
are necessarily in violation of Mather's theorem, the existence of
folded chaotic WGMs with Q-factors larger than $10^{3}$ is a wave
localization effect. This localization is visualized with the help
of Husimi projections which show that the long-lived modes are supported
by a region of the ray phase space that coincides with the heteroclinic/homoclinic
tangles of the unstable periodic orbits whose intensity pattern is
also discernible in the real-space wave plots. In contrast to previous studies of
dynamical localization in chaotic whispering-gallery like cavities, the angular
momentum of the rays is not a slowly diffusing variable for our folded modes;
it is the chaotic manifold structure that allows the wave modes to remain localized
away from all escape windows in phase space.

As seen in the real-space plots, there are varying degrees of emission
into the free space surrounding the structures. In this paper, the
focus has been on the fact that high Q can be preserved in the presence
of waveguide openings, because the latter are more intrusive than
the curvature-induced violation of the total-internal reflection condition
at the dielectric-air interface.

Having characterized the long-lived
states for a range of deformations based on the resonator shape given
in Eq.\ (\ref{eq:ShapeCurve}), explorations of modified shapes are currently in progress, with the
additional goal of further elucidating the Q-factor peaks discussed
in subsection \ref{subsec:Q-factor-oscillations}. By breaking the
reflection symmetry of the cavity and going to higher wavenumbers
in the wave simulations, we expect to improve upon the diagnostic
value of the Husimi projections. We have already performed simulations
for structures without spatial symmetries in order to make contact
with previous work on unidirectional coupling \citep{MicrospiralWG},
but additional work is needed to optimize the Q factors of those shapes.

A complementary approach to the quasibound-state analysis presented here is
to investigate the resonances in transmission or reflection with the
waveguides as input and output. In the stadium as the prototypical
chaotic billiard, the transmission through attached leads has been
studied in \citep{berggrenStadium} in the context of electronic transport,
i.e., with impenetrable walls. Sharp resonances are found to be associated
with all regions of the chaotic phase space, leading to spectral statistics
governed by level repulsion. In our system, the only long-lived modes
are of the folded chaotic WGM type, and their groupings shown
in Fig.\ \ref{fig:Wavenumber-sweeps} are not described by a universal
random-matrix distribution. This goes hand in hand with the observation
that no anti-crossings occur when varying the deformation parameter
$\epsilon$ of Eq.\ (\ref{eq:ShapeCurve}), indicating that all modes
respond to such variations in unison. A study of the transmission
statistics for waveguide-coupled resonators of this type will provide
insight into the interplay of the two types of escape windows in the
phase space depicted in Fig.\ \ref{fig:Poincare-surface-of-section},
where chaotic ray dynamics dominates similarly to the stadium, while
at the same time the mode structure appears to be much simpler provided
that low-Q states are discarded.

\section{Acknowledgment}
This work benefited from access to the University of Oregon high performance
computer, Talapas.

\appendix*

\section{Boundary Integral Implementation}
\subsection{Outline of the method}
To discover and investigate the resonant modes described, we developed
software based on the boundary-integral approach as described in detail
by Heider in Ref.\  \citep{HeiderBIM}. The method is similar to that described in
Ref.\  \citep{Wiersig2003}, which we have also applied to our system.
The latter invariably produces spurious modes that make a large-scale
analysis of the spectrum difficult, whereas the former does not produce any
spurious modes at all (to our knowledge). This significant advantage
can be traced back to the fact that Heider solves a set of simultaneous
disretized of which one half differs from Ref.\  \citep{Wiersig2003} in that
it is obtained by taking an additional normal
derivative of the integral equations for the fields in terms of their values
derivatives on the dielectric interface. The price to be paid for this advantage is
that great care must be taken in in numerically dealing with the singularities
not just in the Green functions but also their derivatives. The following
subsections address these details.

Our software is written using the Julia programming
language, which was chosen for its numeric performance, access to
the necessary special-function libraries with implementations for
complex arguments, and ease of integration with other numerical computing
software. Our implementation is publicly available on GitHub \citep{GitHub}.
It has been written for reusability and allows the user to specify
their own boundary parameterizations. The implementation is written
to utilize multiple processors on a single machine to parallelize
operation where possible and has been tested on OS X and Linux.

To characterize the spectral patterns for folded chaotic WGMs, it
was necessary to scan a wide range of $k$-space. Although the direct
sweep method has the ability to find multiple resonances near an initial
starting guess $k_{0}$, we observed that the ability to discern and
identify relevant starting $k$ values for the secondary RII step
decreases rapidly as $\left\Vert k_{0}-k\right\Vert $ increases.
This is especially true for high-Q modes, as the resonance width for
these modes are very narrow and we empirically observed that the direct
sweep method discovers wider resonances in the vicinity of narrow
resonances more easily which may mask the presence of the narrow resonances.
Therefore, in order to reliably find the high-Q modes it was necessary
to adopt a scanning procedure in which small regions of $k$-space
were examined piecewise and candidate high-Q modes were identified
within those small regions. We used the following algorithm:
\begin{enumerate}
\item \begin{raggedright}
Given a large $k$ range of interest from $k_{\text{min}}$ to $k_{\text{max}}$,
choose a small step size ($k_{\text{step}}=.02$) as a discretization.
\par\end{raggedright}
\item Choose a target $\kappa$ for the imaginary part of $k$. We chose
a small value ($\kappa=.0001$) to target high-Q modes.
\item \begin{raggedright}
Starting at $k_{0}=k_{\text{min}}-i\thinspace\kappa$, execute the
direct sweep procedure to generate candidate resonances.
\par\end{raggedright}
\item \begin{raggedright}
Filter the candidate resonances to keep only those where $k_{0}-k_{\text{step}}\leq k_{\text{cand}}\leq k_{0}+k_{\text{step}}$,
as they will be found again by a closer $k_{0}$ if outside that range
and are less likely to converge to a valid resonance further away
from $k_{0}$.
\par\end{raggedright}
\item \raggedright{}Increment $k_{0}$ by $k_{\text{step}}$ and repeat
this procedure until the desired range has been covered.
\end{enumerate}

Because candidate $k$ values were retained both above and below the
center point $k_{0}$, there was overlap between adjacent center points
and most of the resonances which would eventually be discovered by
the RII procedure would appear twice in the list of candidates. However,
since different starting $k_{0}$ values were used in the direct sweep
procedure, the exact values of these ``duplicate'' $k_{\text{cand}}$
were numerically slightly different as the candidates were discovered
by a linear approximation to the eigenvalue problem around the vicinity
of the initial guess $k_{0}$. In practice, when resolving the more
accurate $k_{f}$ and the boundary field $x_{f}$ using the RII procedure,
these slightly different starting points would converge to the same
resonance within the numerical tolerances specified for the RII algorithm
as well as display an identical spatial field pattern so we concluded
that they were in fact the same mode. Therefore in this system it
would have been possible to be more computationally efficient by combining
or averaging very close candidate $k$ values found in adjacent regions
of $k$-space spanned by the above algorithm, or
in eliminating the overlap and keeping candidates at each step in
the range $k_{0}-k_{\text{step}}/2\leq k_{\text{cand}}\leq k_{0}+k_{\text{step}}/2$
or $k_{0}\leq k_{\text{cand}}\leq k_{0}+k_{\text{step}}$. We chose
to err on the side of the extra computation in the interest of not
missing any important resonances in the scan. It should be noted that
there are other boundary geometries (such as the Reuleaux billiard,
 \citep{PhysRevE.90.022903}) which exhibit very closely spaced doublets
for which a careful approach may be necessary to avoid missing resonances.

Once the list of candidate $k_{0}$ values had been generated, the
final $k_{f}$ and $x_{f}$ values were determined by first filtering
the candidates to exclude very low-Q modes with $\kappa>1.0$ and
then running the direct sweep procedure with starting guess $k_{0}=k_{\text{cand}}$
for each remaining candidate. The reason for the repeated direct sweep
procedure was that since a large number of candidates were initially
generated, we did not store the boundary fields $x$ associated with
them. Since the input to the RII algorithm requires both the starting
$k_{0}$ and the boundary field $x_{0}$, we needed to regenerate
the boundary field. Since the initial candidates $k_{0}$ were already
close to their final $k$, the subsequent direct sweep would in general
produce inputs to the RII procedure that were even closer to the final
convergent values and thus few iterations of the RII loop were required.

\subsection{Some corrections to previous work}

We discovered two errors in the equations listed in the Appendix of
Ref.\  \citep{HeiderBIM}. The first appears in the description of $\tilde{N}$
below equation (A.6). There is a sign error in the last term. The
corrected equation is:

\begin{eqnarray*}
\tilde{N}(t,\tau) & = & \frac{i}{2}\bar{N}(t,\tau)\left\{ (k\,n_{e})^{2}H_{0}^{(1)}(k\,n_{e}\left|x(t)-x(\tau)\right|)-\frac{2k\,n_{e}H_{1}^{(1)}(k\,n_{e}\left|x(t)-x(\tau)\right|)}{\left|x(t)-x(\tau)\right|}\right\} \\
 &  & +\frac{i}{2}\frac{k\,n_{e}x'(t)x'(\tau)}{\left|x(t)-x(\tau)\right|}H_{1}^{(1)}(k\,n_{e}\left|x(t)-x(\tau)\right|)\\
 &  & -\frac{i}{2}\bar{N}(t,\tau)\left\{ (k\,n_{i})^{2}H_{0}^{(1)}(k\,n_{i}\left|x(t)-x(\tau)\right|)-\frac{2k\,n_{i}H_{1}^{(1)}(k\,n_{i}\left|x(t)-x(\tau)\right|)}{\left|x(t)-x(\tau)\right|}\right\} \\
 &  & -\frac{i}{2}\frac{k\,n_{i}x'(t)x'(\tau)}{2\left|x(t)-x(\tau)\right|}H_{1}^{(1)}(k\,n_{i}\left|x(t)-x(\tau)\right|)
\end{eqnarray*}
The other error is in the equation for the diagonal terms $\tilde{N}_{2}(t,t)$,
where the log terms should not use the norm of $x'(t)$ squared, but
simply the norm.The corrected equation is:
\begin{align*}
\tilde{N_{2}}(t,t) & =\frac{\left|x'(t)\right|^{2}}{4\pi}\left[((k\,n_{e})^{2}-(k\,n_{i})^{2})(\pi i-1-2C)\right]\\
 & +\frac{\left|x'(t)\right|^{2}}{4\pi}\left[-2(k\,n_{e})^{2}\text{{ln}}\left(\frac{k\,n_{e}\left|x'(t)\right|}{2}\right)+2(k\,n_{i})^{2}\text{{ln}}\left(\frac{k\,n_{i}\left|x'(t)\right|}{2}\right)\right]
\end{align*}

\subsection{Derivatives of the integral kernels}

Heider's boundary integral approach describes the $A$ matrix and
its use in iteratively finding a resonant solution to the boundary
integral equations. This iterative procedure requires the use of $A'(k)$,
both in the direct sweep procedure (Algorithm 3) and the Newton's
iteration step in the RII procedure (Algorithm 1). The derivation
of the derivative matrix is straightforward and the results are outlined
below, in terms of the constituent equations provided in the paper's
appendix.

The $A'(k)$ matrix is defined as

\[
A'(k)=\begin{bmatrix}-(K_{e}'(k)-K_{i}'(k)) & -(S_{e}'(k)-S_{i}'(k))\\
T_{e}'(k)-T_{i}'(k) & K_{e}^{*}{}'(k)-K_{i}^{*}{}'(k)
\end{bmatrix}
\]

with the $K,S,T,$ and $K^{*}$ integral operators being expressed
after singularity subtraction in terms of equations
\[
\tilde{H}_{1}(k),\tilde{H}_{2}(k),\tilde{H}(k),\tilde{M}_{1}(k),\tilde{M}_{2}(k),\tilde{M}(k),\tilde{N}_{1}(k),\tilde{N}_{2}(k),\tilde{N}(k),\tilde{L}_{1}(k),\tilde{L}_{2}(k),\tilde{L}(k).
\]

All other equations given in the appendix are independent of the wavenumber
$k$.

First the $\tilde{H}$ equations:

\begin{eqnarray*}
\partial_{k}\tilde{H}_{1}(t,\tau,k) & = & -\frac{1}{2\pi}n(\tau)(x(t)-x(\tau))\left[k\,n_{e}^{2}J_{0}(k\,n_{e}\left|x(t)-x(\tau)\right|)-k\,n_{i}^{2}J_{0}(k\,n_{i}\left|x(t)-x(\tau)\right|)\right].\\
\partial_{k}\tilde{H}_{1}(t,t,k) & = & 0\\
\partial_{k}\tilde{H}_{2}(t,\tau,k) & = & \partial_{k}\tilde{H}(t,\tau,k)-\partial_{k}\tilde{H}_{1}(t,\tau,k)\ \text{{ln}}\left(4\,\text{sin}^{2}\frac{t-\tau}{2}\right)\\
\partial_{k}\tilde{H}_{2}(t,\tau,k) & = & \partial\tilde{H}(t,\tau,k)-\partial_{k}\tilde{H}_{1}(t,\tau,k)\ \text{{ln}}\left(4\,\text{sin}^{2}\frac{t-\tau}{2}\right)\\
\partial_{k}\tilde{H}_{2}(t,t,k) & = & 0\\
\partial_{k}\tilde{H}(t,\tau,k) & = & \frac{i}{2}n(\tau)\left(x(t)-x(\tau)\right)\left[k\,n_{e}^{2}H_{0}^{(1)}(k\,n_{e}\left|x(t)-x(\tau)\right|)-k\,n_{i}^{2}H_{0}^{(1)}(k\,n_{i}\left|x(t)-x(\tau)\right|)\right]
\end{eqnarray*}
The $\tilde{H}_{1}^{*}$,$\tilde{H}_{2}^{*}$, and $\tilde{H}^{*}$,
equations used in the starred operators $K^{*}$ are defined in terms
of the unstarred equations with no additional $k$ dependence so their
derivatives can use the results above.

The $\tilde{M}$ equations are:

\begin{eqnarray*}
\partial_{k}\tilde{M}_{1}(t,\tau,k) & = & \frac{\left|x(t)-x(\tau)\right|}{2\pi}\left[n_{e}J_{1}(k\,n_{e}\left|x(t)-x(\tau)\right|)-n_{i}J_{1}(k\,n_{i}\left|x(t)-x(\tau)\right|)\right]\\
\partial_{k}\tilde{M}_{1}(t,t,k) & = & 0\\
\partial_{k}\tilde{M}_{2}(t,\tau,k) & = & \partial_{k}\tilde{M}(t,\tau,k)-\partial_{k}\tilde{M}_{1}(t,\tau,k)\ \text{{ln}}\left(4\,\text{sin}^{2}\frac{t-\tau}{2}\right)\\
\partial_{k}\tilde{M}_{2}(t,t,k) & = & 0\\
\partial_{k}\tilde{M}(t,\tau,k) & = & -\frac{i\left|x(t)-x(\tau)\right|}{2}\left[n_{e}H_{1}^{(1)}(k\,n_{e}\left|x(t)-x(\tau)\right|)-n_{i}H_{1}^{(1)}(k\,n_{i}\left|x(t)-x(\tau)\right|)\right]
\end{eqnarray*}
The $\tilde{N}$ equations are:

\begin{eqnarray*}
\partial_{k}\tilde{N}(t,\tau,k) & = & -\frac{i}{2}\left|x(t)-x(\tau)\right|\bar{N}(t,\tau)\left[k\,n_{e}^{3}H_{1}^{(1)}(k\,n_{e}\left|x(t)-x(\tau)\right|)-k\,n_{i}^{3}H_{1}^{(1)}(k\,n_{i}\left|x(t)-x(\tau)\right|)\right]\\
\qquad\qquad &  & +\frac{i}{2}x'(t)x'(\tau)\left[k\,n_{e}^{2}H_{1}^{(1)}(k\,n_{e}\left|x(t)-x(\tau)\right|)-k\,n_{i}^{2}H_{1}^{(1)}(k\,n_{i}\left|x(t)-x(\tau)\right|)\right]\\
\partial_{k}\tilde{N}_{1}(t,\tau,k) & = & \frac{1}{2\pi}\left|x(t)-x(\tau)\right|\bar{N}(t,\tau)\left[k^{2}n_{e}^{3}J_{1}(k\,n_{e}\left|x(t)-x(\tau)\right|)-k^{2}n_{i}^{3}J_{1}(k\,n_{i}\left|x(t)-x(\tau)\right|)\right]\\
 &  & -\frac{1}{2\pi}x'(t)x'(\tau)\left[k\,n_{e}^{2}J_{0}(k\,n_{e}\left|x(t)-x(\tau)\right|)-k\,n_{i}^{2}J_{0}(k\,n_{i}\left|x(t)-x(\tau)\right|)\right]\\
\partial_{k}\tilde{N}_{1}(t,t,k) & = & \frac{\left|x'(t)\right|^{2}k\left(n_{i}^{2}-n_{e}^{2}\right)}{2\pi}\\
\partial_{k}\tilde{N_{2}}(t,\tau,k) & = & \partial_{k}\tilde{N}(t,\tau,k)-\partial_{k}\tilde{N_{1}}(t,\tau,k)\ \text{{ln}}\left(4\,\text{sin}^{2}\frac{t-\tau}{2}\right)\\
\partial_{k}\tilde{N}_{2}(t,t,k) & = & \frac{\left|x'(t)\right|^{2}}{2\pi}\left[-2k\,n_{e}^{2}\text{\,{ln}\ensuremath{\left(\frac{k\,n_{e}\left|x'(t)\right|}{2}\right)}}+2k\,n_{i}^{2}\text{\,{ln}\ensuremath{\left(\frac{k\,n_{i}\left|x'(t)\right|}{2}\right)}}-k(n_{e}^{2}-n_{i}^{2})(2+2C-\pi i)\right]
\end{eqnarray*}
As in the reference, $C\approx0.5772156649$ refers to the Euler--Mascheroni
constant.

Finally, the $\tilde{L}$ equations:
\begin{eqnarray*}
\partial_{k}\tilde{L}_{1}(t,\tau,k) & = & \frac{\left|x(t)-x(\tau)\right|}{2\pi}\left[(k^{2}n_{e}^{3}J_{1}(k\,n_{e}\left|x(t)-x(\tau)\right|)-k^{2}n_{i}^{3}J_{1}(k\,n_{i}\left|x(t)-x(\tau)\right|))\right]\\
 &  & -\frac{1}{\pi}\left[k\,n_{e}^{2}J_{0}(k\,n_{e}\left|x(t)-x(\tau)\right|))-k\,n_{i}^{2}J_{0}(k\,n_{i}\left|x(t)-x(\tau)\right|))\right]\\
\partial_{k}\tilde{L}_{1}(t,t,k) & = & \frac{k(n_{i}^{2}-n_{e}^{2})}{\pi}\\
\partial_{k}\tilde{L}_{2}(t,\tau,k) & = & \partial_{k}\tilde{L}(t,\tau,k)-\partial_{k}\tilde{L}_{1}(t,\tau,k)\ \text{{ln}}\left(4\,\text{sin}^{2}\frac{t-\tau}{2}\right)\\
\partial_{k}\tilde{L}_{2}(t,t,k) & = & -\frac{1}{\pi}\left[2k\,n_{e}^{2}\text{\,{ln}}(k\,n_{e})-2k\,n_{i}^{2}\text{\,{ln}}(k\,n_{i})+k\left(n_{e}^{2}-n_{i}^{2}\right)\left(2\text{\,{ln}}(\frac{\left|x'(t)\right|}{2})-\pi i+2C+1\right)\right]\\
\partial_{k}\tilde{L}(t,\tau,k) & = & -\frac{i}{2}\left|x(t)-x(\tau)\right|\left[k^{2}n_{e}^{3}H_{1}^{(1)}(k\,n_{e}\left|x(t)-x(\tau)\right|)-k^{2}n_{i}^{3}H_{1}^{(1)}(k\,n_{i}\left|x(t)-x(\tau)\right|)\right]\\
 &  & +i\left[k\,n_{e}^{2}H_{0}^{(1)}(k\,n_{e}\left|x(t)-x(\tau)\right|)-k\,n_{i}^{2}H_{0}^{(1)}(k\,n_{i}\left|x(t)-x(\tau)\right|)\right]
\end{eqnarray*}
It is hoped that the explicit formulas given here provide a complete
picture of the computational method we employed in finding the quasibound
states.

\bibliographystyle{apsrev}

%\bibliography{FoldedWGmodes}

\begin{thebibliography}{38}
\expandafter\ifx\csname natexlab\endcsname\relax\def\natexlab#1{#1}\fi
\expandafter\ifx\csname bibnamefont\endcsname\relax
  \def\bibnamefont#1{#1}\fi
\expandafter\ifx\csname bibfnamefont\endcsname\relax
  \def\bibfnamefont#1{#1}\fi
\expandafter\ifx\csname citenamefont\endcsname\relax
  \def\citenamefont#1{#1}\fi
\expandafter\ifx\csname url\endcsname\relax
  \def\url#1{}\fi
\expandafter\ifx\csname urlprefix\endcsname\relax\def\urlprefix{}\fi
\providecommand{\bibinfo}[2]{#2}
\providecommand{\eprint}[2][]{\url{#2}}

\bibitem[{\citenamefont{Cao and Wiersig}(2015)}]{CaoWiersigReview}
\bibinfo{author}{\bibfnamefont{H.}~\bibnamefont{Cao}} \bibnamefont{and}
  \bibinfo{author}{\bibfnamefont{J.}~\bibnamefont{Wiersig}},
  \href{https://link.aps.org/doi/10.1103/RevModPhys.87.61}{
  \bibinfo{journal}{Reviews of Modern Physics} \textbf{\bibinfo{volume}{87}},
  \bibinfo{pages}{61} (\bibinfo{year}{2015})}.

\bibitem[{\citenamefont{N{\"o}ckel and Stone}(1997)}]{Nockel:1997lr}
\bibinfo{author}{\bibfnamefont{J.~U.} \bibnamefont{N{\"o}ckel}}
  \bibnamefont{and} \bibinfo{author}{\bibfnamefont{A.~D.} \bibnamefont{Stone}},
  \href{http://dx.doi.org/10.1038/385045a0}
  {\bibinfo{journal}{Nature} \textbf{\bibinfo{volume}{385}}, \bibinfo{pages}{45}
  (\bibinfo{year}{1997})}.

\bibitem[{\citenamefont{Liu et~al.}(2013)\citenamefont{Liu, Jiang, Li, Xiao,
  Wang, Ren, Zhang, Yang, and Gong}}]{ISI:000320621600008}
\bibinfo{author}{\bibfnamefont{Z.-P.} \bibnamefont{Liu}},
  \bibinfo{author}{\bibfnamefont{X.-F.} \bibnamefont{Jiang}},
  \bibinfo{author}{\bibfnamefont{Y.}~\bibnamefont{Li}},
  \bibinfo{author}{\bibfnamefont{Y.-F.} \bibnamefont{Xiao}},
  \bibinfo{author}{\bibfnamefont{L.}~\bibnamefont{Wang}},
  \bibinfo{author}{\bibfnamefont{J.-L.} \bibnamefont{Ren}},
  \bibinfo{author}{\bibfnamefont{S.-J.} \bibnamefont{Zhang}},
  \bibinfo{author}{\bibfnamefont{H.}~\bibnamefont{Yang}}, \bibnamefont{and}
  \bibinfo{author}{\bibfnamefont{Q.}~\bibnamefont{Gong}},
  \href{http://dx.doi.org/10.1063/1.4809724}{
  \bibinfo{journal}{Applied Physics Letters} \textbf{\bibinfo{volume}{102}},
  \bibinfo{pages}{221108}
  (\bibinfo{year}{2013})}.

\bibitem[{\citenamefont{Lazutkin}(1993)}]{LazutkinBook}
\bibinfo{author}{\bibfnamefont{V.~F.} \bibnamefont{Lazutkin}},
  \emph{\bibinfo{title}{KAM Theory and Semiclassical Approximations to
  Eigenfunctions}} (\bibinfo{publisher}{Springer}, \bibinfo{address}{New York},
  \bibinfo{year}{1993}).

\bibitem[{\citenamefont{Schwieters et~al.}(1996)\citenamefont{Schwieters,
  Alford, and Delos}}]{PhysRevB.54.10652}
\bibinfo{author}{\bibfnamefont{C.~D.} \bibnamefont{Schwieters}},
  \bibinfo{author}{\bibfnamefont{J.~A.} \bibnamefont{Alford}},
  \bibnamefont{and} \bibinfo{author}{\bibfnamefont{J.~B.} \bibnamefont{Delos}},
  \href{https://link.aps.org/doi/10.1103/PhysRevB.54.10652}{
  \bibinfo{journal}{Phys. Rev. B} \textbf{\bibinfo{volume}{54}},
  \bibinfo{pages}{10652} (\bibinfo{year}{1996})}.

\bibitem[{\citenamefont{Mather}(1984)}]{MatherInvariant}
\bibinfo{author}{\bibfnamefont{J.~N.} \bibnamefont{Mather}},
\href{http://dx.doi.org/10.1017/S0143385700002455}{
\bibinfo{journal}{Ergodic Theory and Dynamical Systems} \textbf{\bibinfo{volume}{4}},
\bibinfo{pages}{301} (\bibinfo{year}{1984})}.

\bibitem[{\citenamefont{Bunimovich}(1979)}]{BunimovichStadium}
\bibinfo{author}{\bibfnamefont{L.~A.} \bibnamefont{Bunimovich}},
  \href{https://doi.org/10.1007/BF01197884}{
  \bibinfo{journal}{Commun. Math. Phys.}
  \textbf{\bibinfo{volume}{65}}, \bibinfo{pages}{295} (\bibinfo{year}{1979})}.

\bibitem[{\citenamefont{N{\"o}ckel et~al.}(1994)\citenamefont{N{\"o}ckel,
  Stone, and Chang}}]{Nockel:94}
\bibinfo{author}{\bibfnamefont{J.~U.} \bibnamefont{N{\"o}ckel}},
  \bibinfo{author}{\bibfnamefont{A.~D.} \bibnamefont{Stone}}, \bibnamefont{and}
  \bibinfo{author}{\bibfnamefont{R.~K.} \bibnamefont{Chang}},
  \href{http://ol.osa.org/abstract.cfm?URI=ol-19-21-1693}{
  \bibinfo{journal}{Opt. Lett.} \textbf{\bibinfo{volume}{19}},
  \bibinfo{pages}{1693} (\bibinfo{year}{1994})}.

\bibitem[{\citenamefont{Backes et~al.}(1998)\citenamefont{Backes, Cleaver,
  Heberle, and K{\"o}hler}}]{backes:VerticalAndNotches}
\bibinfo{author}{\bibfnamefont{S.~A.} \bibnamefont{Backes}},
  \bibinfo{author}{\bibfnamefont{J.~R.~A.} \bibnamefont{Cleaver}},
  \bibinfo{author}{\bibfnamefont{A.~P.} \bibnamefont{Heberle}},
  \bibnamefont{and} \bibinfo{author}{\bibfnamefont{K.}~\bibnamefont{K{\"o}hler}},
  \href{https://avs.scitation.org/doi/abs/10.1116/1.590415}{
  \bibinfo{journal}{J. Vac. Sci. Technol. B}
  \textbf{\bibinfo{volume}{16}}, \bibinfo{pages}{3817} (\bibinfo{year}{1998})}.

\bibitem[{\citenamefont{Lv et~al.}(2013)\citenamefont{Lv, Huang, Yang, Long,
  Zou, Yao, Jin, Xiao, and Du}}]{VerticalLossWithWG}
\bibinfo{author}{\bibfnamefont{X.-M.} \bibnamefont{Lv}},
  \bibinfo{author}{\bibfnamefont{Y.-Z.} \bibnamefont{Huang}},
  \bibinfo{author}{\bibfnamefont{Y.-D.} \bibnamefont{Yang}},
  \bibinfo{author}{\bibfnamefont{H.}~\bibnamefont{Long}},
  \bibinfo{author}{\bibfnamefont{L.-X.} \bibnamefont{Zou}},
  \bibinfo{author}{\bibfnamefont{Q.-F.} \bibnamefont{Yao}},
  \bibinfo{author}{\bibfnamefont{X.}~\bibnamefont{Jin}},
  \bibinfo{author}{\bibfnamefont{J.-L.} \bibnamefont{Xiao}}, \bibnamefont{and}
  \bibinfo{author}{\bibfnamefont{Y.}~\bibnamefont{Du}},
  \href{http://www.opticsexpress.org/abstract.cfm?URI=oe-21-13-16069}{
  \bibinfo{journal}{Opt.
  Express} \textbf{\bibinfo{volume}{21}}, \bibinfo{pages}{16069}
  (\bibinfo{year}{2013})}.

\bibitem[{\citenamefont{Li and Liu}(1997)}]{VerticalSpreadingDisk}
\bibinfo{author}{\bibfnamefont{B.-J.} \bibnamefont{Li}} \bibnamefont{and}
  \bibinfo{author}{\bibfnamefont{P.-L.} \bibnamefont{Liu}},
  \href{http://dx.doi.org/10.1109/3.622627}{
  \bibinfo{journal}{IEEE J. Quantum Electron.}
  \textbf{\bibinfo{volume}{33}}, \bibinfo{pages}{1489} (\bibinfo{year}{1997})}.

\bibitem[{\citenamefont{Yan et~al.}(2014)\citenamefont{Yan, Shi, Li, Li, and
  Zhang}}]{LimaconDirectionality}
\bibinfo{author}{\bibfnamefont{C.}~\bibnamefont{Yan}},
  \bibinfo{author}{\bibfnamefont{J.}~\bibnamefont{Shi}},
  \bibinfo{author}{\bibfnamefont{P.}~\bibnamefont{Li}},
  \bibinfo{author}{\bibfnamefont{H.}~\bibnamefont{Li}}, \bibnamefont{and}
  \bibinfo{author}{\bibfnamefont{J.}~\bibnamefont{Zhang}},
  \href{http://doi.org/101016/j.optlastec.2013.08.009}
  {\bibinfo{journal}{Opt. Laser Technol.} \textbf{\bibinfo{volume}{56}},
  \bibinfo{pages}{285} (\bibinfo{year}{2014})}.

\bibitem[{\citenamefont{Song et~al.}(2012)\citenamefont{Song, Ge, Redding, and
  Cao}}]{ChaoticChannelingPRL}
\bibinfo{author}{\bibfnamefont{Q.}~\bibnamefont{Song}},
  \bibinfo{author}{\bibfnamefont{L.}~\bibnamefont{Ge}},
  \bibinfo{author}{\bibfnamefont{B.}~\bibnamefont{Redding}}, \bibnamefont{and}
  \bibinfo{author}{\bibfnamefont{H.}~\bibnamefont{Cao}},
  \href{https://doi.org/10.1103/PhysRevLett.108.243902}{
  \bibinfo{journal}{Phys. Rev. Lett.} \textbf{\bibinfo{volume}{108}}, \bibinfo{pages}{243902}
  (\bibinfo{year}{2012})}.

\bibitem[{\citenamefont{Liu et~al.}(2018)\citenamefont{Liu, Sun, Wang, Yu, Xu,
  Huang, Xiao, and Song}}]{EndFireInjection}
\bibinfo{author}{\bibfnamefont{S.}~\bibnamefont{Liu}},
  \bibinfo{author}{\bibfnamefont{W.}~\bibnamefont{Sun}},
  \bibinfo{author}{\bibfnamefont{Y.}~\bibnamefont{Wang}},
  \bibinfo{author}{\bibfnamefont{X.}~\bibnamefont{Yu}},
  \bibinfo{author}{\bibfnamefont{K.}~\bibnamefont{Xu}},
  \bibinfo{author}{\bibfnamefont{Y.}~\bibnamefont{Huang}},
  \bibinfo{author}{\bibfnamefont{S.}~\bibnamefont{Xiao}}, \bibnamefont{and}
  \bibinfo{author}{\bibfnamefont{Q.}~\bibnamefont{Song}},
  \href{http://www.osapublishing.org/optica/abstract.cfm?URI=optica-5-5-612}{
  \bibinfo{journal}{Optica} \textbf{\bibinfo{volume}{5}}, \bibinfo{pages}{612}
  (\bibinfo{year}{2018})}.

\bibitem[{\citenamefont{Yang et~al.}(2014)\citenamefont{Yang, Zhang, Huang, and
  Poon}}]{MicrospiralWG}
\bibinfo{author}{\bibfnamefont{Y.-D.} \bibnamefont{Yang}},
  \bibinfo{author}{\bibfnamefont{Y.}~\bibnamefont{Zhang}},
  \bibinfo{author}{\bibfnamefont{Y.-Z.} \bibnamefont{Huang}}, \bibnamefont{and}
  \bibinfo{author}{\bibfnamefont{A.~W.} \bibnamefont{Poon}},
  \href{https://doi.org/10.1364/OE.22.000824}
  {\bibinfo{journal}{Opt. Express} \textbf{\bibinfo{volume}{22}},
  \bibinfo{pages}{824} (\bibinfo{year}{2014})}.

\bibitem[{\citenamefont{Che et~al.}(2010)\citenamefont{Che, Lin, Huang, Yang,
  Xiao, and Du}}]{SquareMicrolasers}
\bibinfo{author}{\bibfnamefont{K.-J.} \bibnamefont{Che}},
  \bibinfo{author}{\bibfnamefont{J.-D.} \bibnamefont{Lin}},
  \bibinfo{author}{\bibfnamefont{Y.-Z.} \bibnamefont{Huang}},
  \bibinfo{author}{\bibfnamefont{Y.-D.} \bibnamefont{Yang}},
  \bibinfo{author}{\bibfnamefont{J.-L.} \bibnamefont{Xiao}}, \bibnamefont{and}
  \bibinfo{author}{\bibfnamefont{Y.}~\bibnamefont{Du}},
  \href{http://doi.org/10.1109/LPT.2010.2057417}
  {\bibinfo{journal}{IEEE
  Photon. Technol. Lett.} \textbf{\bibinfo{volume}{22}},
  \bibinfo{pages}{1370} (\bibinfo{year}{2010})}.

\bibitem[{\citenamefont{YongZhen et~al.}(2009)\citenamefont{Y.-Z. Huang,
Y.-D. Yang, S.-J. Wang, J.-L. Xiao, K.-J. Che and Y. Du}}]{TriangleAndSquareWG}
\bibinfo{author}{\bibfnamefont{Y.-Z.}~\bibnamefont{Huang}},
  \bibinfo{author}{\bibfnamefont{Y.-D.}~\bibnamefont{Yang}},
  \bibinfo{author}{\bibfnamefont{S.-J.}~\bibnamefont{Wang}},
  \bibinfo{author}{\bibfnamefont{J.-L.}~\bibnamefont{Xiao}},
  \bibinfo{author}{\bibfnamefont{K.-J.}~\bibnamefont{Che}}, \bibnamefont{and}
  \bibinfo{author}{\bibfnamefont{Y.}~\bibnamefont{Du}},
  \href{https://doi.org/10.1007/s11431-009-0306-y}
  {\bibinfo{journal}{Sci. China Ser. E: Technol. Sci.}
  \textbf{\bibinfo{volume}{52}}, \bibinfo{pages}{3447} (\bibinfo{year}{2009})}.

\bibitem[{\citenamefont{Zou et~al.}(2013)\citenamefont{Zou, Lv, Huang, Long,
  Xiao, Yao, Lin, and Du}}]{octagonWithLead}
\bibinfo{author}{\bibfnamefont{L.}~\bibnamefont{Zou}},
  \bibinfo{author}{\bibfnamefont{X.}~\bibnamefont{Lv}},
  \bibinfo{author}{\bibfnamefont{Y.}~\bibnamefont{Huang}},
  \bibinfo{author}{\bibfnamefont{H.}~\bibnamefont{Long}},
  \bibinfo{author}{\bibfnamefont{J.}~\bibnamefont{Xiao}},
  \bibinfo{author}{\bibfnamefont{Q.}~\bibnamefont{Yao}},
  \bibinfo{author}{\bibfnamefont{J.}~\bibnamefont{Lin}}, \bibnamefont{and}
  \bibinfo{author}{\bibfnamefont{Y.}~\bibnamefont{Du}},
  \href{http://doi.org/10.1109/JSTQE.2013.2244566}
  {\bibinfo{journal}{IEEE
  J. Sel. Top. in Quantum Electron.}
  \textbf{\bibinfo{volume}{19}}, \bibinfo{pages}{1501808}
  (\bibinfo{year}{2013})}.

\bibitem[{\citenamefont{Fukushima et~al.}(2007)\citenamefont{Fukushima, Tanaka,
  and Harayama}}]{Fukushima:07}
\bibinfo{author}{\bibfnamefont{T.}~\bibnamefont{Fukushima}},
  \bibinfo{author}{\bibfnamefont{T.}~\bibnamefont{Tanaka}}, \bibnamefont{and}
  \bibinfo{author}{\bibfnamefont{T.}~\bibnamefont{Harayama}},
  \href{{http://ol.osa.org/abstract.cfm?URI=ol-32-23-3397}}
  {\bibinfo{journal}{Opt. Lett.} \textbf{\bibinfo{volume}{32}},
  \bibinfo{pages}{3397} (\bibinfo{year}{2007})}.

\bibitem[{\citenamefont{Prange et~al.}(2001)\citenamefont{Prange, Narevich, and
  Zaitsev}}]{Prange2001}
\bibinfo{author}{\bibfnamefont{R.~E.} \bibnamefont{Prange}},
  \bibinfo{author}{\bibfnamefont{R.}~\bibnamefont{Narevich}}, \bibnamefont{and}
  \bibinfo{author}{\bibfnamefont{O.}~\bibnamefont{Zaitsev}},
  \href{{http://stacks.iop.org/1402-4896/T90/134}}
  {\bibinfo{journal}{Phys. Scr.} \textbf{\bibinfo{volume}{T90}},
  \bibinfo{pages}{134} (\bibinfo{year}{2001})}.

\bibitem[{\citenamefont{Gmachl et~al.}(1998)\citenamefont{Gmachl, Capasso,
  Narimanov, N{\"o}ckel, Stone, Faist, Sivco, and Cho}}]{Gmachl05061998}
\bibinfo{author}{\bibfnamefont{C.}~\bibnamefont{Gmachl}},
  \bibinfo{author}{\bibfnamefont{F.}~\bibnamefont{Capasso}},
  \bibinfo{author}{\bibfnamefont{E.~E.} \bibnamefont{Narimanov}},
  \bibinfo{author}{\bibfnamefont{J.~U.} \bibnamefont{N{\"o}ckel}},
  \bibinfo{author}{\bibfnamefont{A.~D.} \bibnamefont{Stone}},
  \bibinfo{author}{\bibfnamefont{J.}~\bibnamefont{Faist}},
  \bibinfo{author}{\bibfnamefont{D.~L.} \bibnamefont{Sivco}}, \bibnamefont{and}
  \bibinfo{author}{\bibfnamefont{A.~Y.} \bibnamefont{Cho}},
  \href{http://www.sciencemag.org/content/280/5369/1556.abstract}
  {\bibinfo{journal}{Science} \textbf{\bibinfo{volume}{280}},
  \bibinfo{pages}{1556} (\bibinfo{year}{1998})}.

  \bibitem[{\citenamefont{Babic and Buldyrev}(1972)}]{BabicBook1972}
  \bibinfo{author}{\bibfnamefont{V.~M.} \bibnamefont{Babic}} \bibnamefont{and}
    \bibinfo{author}{\bibfnamefont{V.~S.} \bibnamefont{Buldyrev}},
    \emph{\bibinfo{title}{Short-wavelength diffraction theory}}
    (\bibinfo{publisher}{Springer Verlag}, \bibinfo{address}{Berlin},
    \bibinfo{year}{1972}).


\bibitem[{\citenamefont{Fang et~al.}(2005)\citenamefont{Fang, Cao, Podolskiy,
  and Narimanov}}]{FangCao2005}
\bibinfo{author}{\bibfnamefont{W.}~\bibnamefont{Fang}},
  \bibinfo{author}{\bibfnamefont{H.}~\bibnamefont{Cao}},
  \bibinfo{author}{\bibfnamefont{V.~A.} \bibnamefont{Podolskiy}},
  \bibinfo{author}{\bibfnamefont{E.~E.} \bibnamefont{Narimanov}},
  \href{http://www.opticsexpress.org/abstract.cfm?URI=oe-13-15-5641}{
  \bibinfo{journal}{Opt. Express} \textbf{\bibinfo{volume}{13}},
  \bibinfo{pages}{5641} (\bibinfo{year}{2005})}.

\bibitem[{\citenamefont{N{\"o}ckel}(2001)\citenamefont{N{\"o}ckel}}]{Noeckel:T90-2001}
\bibinfo{author}{\bibfnamefont{J.~U.}~\bibnamefont{N{\"o}ckel}},
  \href{http://stacks.iop.org/1402-4896/T90/263}{\bibinfo{journal}{Phys. Scr} \textbf{\bibinfo{volume}{T90}},
  \bibinfo{pages}{263} (\bibinfo{year}{2001})}.

\bibitem[{\citenamefont{Lackner et~al.}(2013)\citenamefont{Lackner, Brezinova,
  Burgdorfer, and Libisch}}]{ISI:000323333000016}
\bibinfo{author}{\bibfnamefont{F.}~\bibnamefont{Lackner}},
  \bibinfo{author}{\bibfnamefont{I.}~\bibnamefont{Brezinova}},
  \bibinfo{author}{\bibfnamefont{J.}~\bibnamefont{Burgd{\"o}rfer}},
  \bibnamefont{and} \bibinfo{author}{\bibfnamefont{F.}~\bibnamefont{Libisch}},
  \href{https://doi.org/10.1103/PhysRevE.88.022916}{
  \bibinfo{journal}{Phys. Rev. E} \textbf{\bibinfo{volume}{88}}, \bibinfo{pages}{022916}
  (\bibinfo{year}{2013})}.

  \bibitem[{\citenamefont{{Fern{\'a}ndez Guasti}}(1992)}]{Squircle}
  \bibinfo{author}{\bibfnamefont{M.}~\bibnamefont{{Fern{\'a}ndez Guasti}}},
    \href{https://doi.org/10.1080/0020739920230607}{
    \bibinfo{journal}{Int. J. Math. Educ. Sci.
  Technol.} \textbf{\bibinfo{volume}{23}}, \bibinfo{pages}{895}
    (\bibinfo{year}{1992})}.

\bibitem[{\citenamefont{Heider}(2010)}]{HeiderBIM}
\bibinfo{author}{\bibfnamefont{P.}~\bibnamefont{Heider}},
  \href{https://doi.org/10.1016/j.camwa.2010.06.044}
  {\bibinfo{journal}{Comput. Math. Appl.}
  \textbf{\bibinfo{volume}{60}}, \bibinfo{pages}{1620} (\bibinfo{year}{2010})}.

\bibitem[{\citenamefont{N{\"o}ckel and Stone}(1996)}]{NockelMcBook1}
\bibinfo{author}{\bibfnamefont{J.~U.} \bibnamefont{N{\"o}ckel}}
  \bibnamefont{and} \bibinfo{author}{\bibfnamefont{A.~D.} \bibnamefont{Stone}},
  in \href{https://arxiv.org/abs/physics/0203063}{
  \emph{\bibinfo{booktitle}{Optical Processes in Microcavities}}, edited by
  \bibinfo{editor}{\bibfnamefont{R.~K.} \bibnamefont{Chang}} \bibnamefont{and}
  \bibinfo{editor}{\bibfnamefont{A.~J.} \bibnamefont{Campillo}}
  (\bibinfo{publisher}{World Scientific, Singapore}, \bibinfo{year}{1996})}.

\bibitem[{\citenamefont{Mandelshtam}(2001)}]{Mandelshtam:2001fk}
\bibinfo{author}{\bibfnamefont{V.~A.} \bibnamefont{Mandelshtam}},
  \href{http://www.sciencedirect.com/science/article/pii/S0079656500000327}
  {\bibinfo{journal}{Prog. Nucl. Magn. Reson. Spectrosc.}
  \textbf{\bibinfo{volume}{38}}, \bibinfo{pages}{159} (\bibinfo{year}{2001})}.

\bibitem[{\citenamefont{Oskooi et~al.}(2010)\citenamefont{Oskooi, Roundy,
  Ibanescu, Bermel, Joannopoulos, and Johnson}}]{MEEP-citation}
\bibinfo{author}{\bibfnamefont{A.~F.} \bibnamefont{Oskooi}},
  \bibinfo{author}{\bibfnamefont{D.}~\bibnamefont{Roundy}},
  \bibinfo{author}{\bibfnamefont{M.}~\bibnamefont{Ibanescu}},
  \bibinfo{author}{\bibfnamefont{P.}~\bibnamefont{Bermel}},
  \bibinfo{author}{\bibfnamefont{J.~D.} \bibnamefont{Joannopoulos}},
  \bibnamefont{and} \bibinfo{author}{\bibfnamefont{S.~G.}
  \bibnamefont{Johnson}},
  \href{http://www.sciencedirect.com/science/article/pii/S001046550900383X}{
  \bibinfo{journal}{Comput. Phys. Commun.}
  \textbf{\bibinfo{volume}{181}}, \bibinfo{pages}{687} (\bibinfo{year}{2010})}.

\bibitem[{\citenamefont{Lichtenberg and Lieberman}(1992)}]{LichtenbergBook1992}
\bibinfo{author}{\bibfnamefont{A.~J.} \bibnamefont{Lichtenberg}}
  \bibnamefont{and} \bibinfo{author}{\bibfnamefont{M.~A.}
  \bibnamefont{Lieberman}}, \emph{\bibinfo{title}{Regular and Chaotic Dynamics,
  Second Edition}} (\bibinfo{publisher}{Springer-Verlag}, \bibinfo{address}{New
  York}, \bibinfo{year}{1992}).

\bibitem[{\citenamefont{N{\"o}ckel and Chang}(2002)}]{NoeckelChapter2002}
\bibinfo{author}{\bibfnamefont{J.~U.} \bibnamefont{N{\"o}ckel}}
  \bibnamefont{and} \bibinfo{author}{\bibfnamefont{R.~K.} \bibnamefont{Chang}},
  in \href{https://arxiv.org/abs/physics/0406134}{
  \emph{\bibinfo{booktitle}{Cavity-Enhanced Spectroscopies}}, edited by
  \bibinfo{editor}{\bibfnamefont{R.~D.} \bibnamefont{van Zee}}
  \bibnamefont{and} \bibinfo{editor}{\bibfnamefont{J.~P.} \bibnamefont{Looney}}
  (\bibinfo{publisher}{Academic Press, San Diego}, \bibinfo{year}{2002})},
  vol.~\bibinfo{volume}{40} of \emph{\bibinfo{series}{Experimental Methods in
  the Physical Sciences}}, pp. \bibinfo{pages}{185--226}.

\bibitem[{\citenamefont{Dietz et~al.}(2014)\citenamefont{Dietz, Guhr, Gutkin,
  Miski-Oglu, and Richter}}]{PhysRevE.90.022903}
\bibinfo{author}{\bibfnamefont{B.}~\bibnamefont{Dietz}},
  \bibinfo{author}{\bibfnamefont{T.}~\bibnamefont{Guhr}},
  \bibinfo{author}{\bibfnamefont{B.}~\bibnamefont{Gutkin}},
  \bibinfo{author}{\bibfnamefont{M.}~\bibnamefont{Miski-Oglu}},
  \bibnamefont{and} \bibinfo{author}{\bibfnamefont{A.}~\bibnamefont{Richter}},
  \href{https://link.aps.org/doi/10.1103/PhysRevE.90.022903}{
  \bibinfo{journal}{Phys. Rev. E} \textbf{\bibinfo{volume}{90}},
  \bibinfo{pages}{022903} (\bibinfo{year}{2014})}.

\bibitem[{\citenamefont{Wiersig}(2006)}]{WiersigAnticrossingModes}
\bibinfo{author}{\bibfnamefont{J.}~\bibnamefont{Wiersig}},
  \href{https://link.aps.org/doi/10.1103/PhysRevLett.97.253901}{
  \bibinfo{journal}{Phys. Rev. Lett.} \textbf{\bibinfo{volume}{97}},
  \bibinfo{pages}{253901} (\bibinfo{year}{2006})}.

\bibitem[{\citenamefont{Foster et~al.}(2007)\citenamefont{Foster, Cook, and
  N{\"o}ckel}}]{Foster:07}
\bibinfo{author}{\bibfnamefont{D.~H.} \bibnamefont{Foster}},
  \bibinfo{author}{\bibfnamefont{A.~K.} \bibnamefont{Cook}}, \bibnamefont{and}
  \bibinfo{author}{\bibfnamefont{J.~U.} \bibnamefont{N{\"o}ckel}},
  \href{https://doi.org/10.1364/OL.32.001764}{
  \bibinfo{journal}{Opt. Lett.} \textbf{\bibinfo{volume}{32}},
  \bibinfo{pages}{1764} (\bibinfo{year}{2007})}.

\bibitem[{\citenamefont{Unterhinninghofen
  et~al.}(2008)\citenamefont{Unterhinninghofen, Wiersig, and
  Hentschel}}]{Unterinninghofen2008}
\bibinfo{author}{\bibfnamefont{J.}~\bibnamefont{Unterhinninghofen}},
  \bibinfo{author}{\bibfnamefont{J.}~\bibnamefont{Wiersig}}, \bibnamefont{and}
  \bibinfo{author}{\bibfnamefont{M.}~\bibnamefont{Hentschel}},
  \href{https://link.aps.org/doi/10.1103/PhysRevE.78.016201}{
  \bibinfo{journal}{Phys. Rev. E} \textbf{\bibinfo{volume}{78}},
  \bibinfo{pages}{016201} (\bibinfo{year}{2008})}.

\bibitem[{\citenamefont{Hackenbroich and N{\"o}ckel}(1997)}]{Hackenbroich97}
\bibinfo{author}{\bibfnamefont{G.}~\bibnamefont{Hackenbroich}}
  \bibnamefont{and} \bibinfo{author}{\bibfnamefont{J.~U.}
  \bibnamefont{N{\"o}ckel}},
  \href{http://stacks.iop.org/0295-5075/39/371}{
  \bibinfo{journal}{Europhys. Lett.}
  \textbf{\bibinfo{volume}{39}}, \bibinfo{pages}{371} (\bibinfo{year}{1997})}.

\bibitem[{\citenamefont{Brack and Bhaduri}(2003)}]{BrackBook1997}
\bibinfo{author}{\bibfnamefont{M.}~\bibnamefont{Brack}} \bibnamefont{and}
  \bibinfo{author}{\bibfnamefont{R.~K.} \bibnamefont{Bhaduri}},
  \emph{\bibinfo{title}{Semiclassical Physics}} (\bibinfo{publisher}{Westview
  Press}, \bibinfo{address}{Boulder, Co.}, \bibinfo{year}{2003}).

\bibitem[{\citenamefont{Brack et~al.}(1993)\citenamefont{Brack, Bhaduri, Law,
  and Murthy}}]{PhysRevLett.70.568}
\bibinfo{author}{\bibfnamefont{M.}~\bibnamefont{Brack}},
  \bibinfo{author}{\bibfnamefont{R.~K.} \bibnamefont{Bhaduri}},
  \bibinfo{author}{\bibfnamefont{J.}~\bibnamefont{Law}}, \bibnamefont{and}
  \bibinfo{author}{\bibfnamefont{M.~V.~N.} \bibnamefont{Murthy}},
  \href{https://link.aps.org/doi/10.1103/PhysRevLett.70.568}{
  \bibinfo{journal}{Phys. Rev. Lett.} \textbf{\bibinfo{volume}{70}},
  \bibinfo{pages}{568} (\bibinfo{year}{1993})}.

\bibitem[{\citenamefont{Ishio et~al.}(2001)\citenamefont{Ishio, Saichev,
  Sadreev, and Berggren}}]{berggrenStadium}
\bibinfo{author}{\bibfnamefont{H.}~\bibnamefont{Ishio}},
  \bibinfo{author}{\bibfnamefont{A.~I.} \bibnamefont{Saichev}},
  \bibinfo{author}{\bibfnamefont{A.~F.} \bibnamefont{Sadreev}},
  \bibnamefont{and} \bibinfo{author}{\bibfnamefont{K.-F.}
  \bibnamefont{Berggren}},
  \href{http://www.sciencedirect.com/science/article/pii/S0010465501003204}{
  \bibinfo{journal}{Comput. Phys. Commun.}
  \textbf{\bibinfo{volume}{142}}, \bibinfo{pages}{64} (\bibinfo{year}{2001})}.

\bibitem[{\citenamefont{Wiersig}(2003)}]{Wiersig2003}
\bibinfo{author}{\bibfnamefont{J.}~\bibnamefont{Wiersig}},
  \href{http://stacks.iop.org/1464-4258/5/53}{
  \bibinfo{journal}{J. Opt. A: Pure Appl. Opt.}
  \textbf{\bibinfo{volume}{5}}, \bibinfo{pages}{53} (\bibinfo{year}{2003})}.

  \bibitem{GitHub}
  \url{https://github.com/kahliburke/BoundaryIntegralMethod.jl}

\end{thebibliography}

\end{document}